\documentclass[]{aastex631}

\newcommand{\ka}{K$\alpha$~}

\usepackage{amsmath,bm}
\usepackage{longtable}
\usepackage{natbib}

\defcitealias{trindadefalcao2023a}{Paper 1}
\usepackage[T1]{fontenc}
\usepackage{newtxtext,newtxmath}
\DeclareRobustCommand{\VAN}[3]{#2}
\let\VANthebibliography\thebibliography
\def\thebibliography{\DeclareRobustCommand{\VAN}[3]{##3}\VANthebibliography}

\makeatletter
\DeclareRobustCommand{\element}[1]{\@element#1\@nil}
\def\@element#1#2\@nil{%
  #1%
  \if\relax#2\relax\else\MakeLowercase{#2}\fi}
\makeatother

\usepackage{graphicx}	
\usepackage{bm}
\usepackage{amsmath}

\usepackage{booktabs}
\usepackage{gensymb}

\begin{document}


\title[NGC 5728]{Discovery of kiloparsec-scale semi-relativistic Fe K$\alpha$ complex emission in NGC 5728}

\correspondingauthor{Anna Trindade Falcao}
\email{anna.trindade\_falcao@cfa.harvard.edu}

\author{Anna Trindade Falcao}
\affiliation{Harvard-Smithsonian Center for Astrophysics, \\
60 Garden St., Cambridge, MA 02138, USA}

\author{G. Fabbiano}
\affiliation{Harvard-Smithsonian Center for Astrophysics, \\
60 Garden St., Cambridge, MA 02138, USA}

\author{M. Elvis}
\affiliation{Harvard-Smithsonian Center for Astrophysics, \\
60 Garden St., Cambridge, MA 02138, USA}

\author{A. Paggi}
\affiliation{Dipartimento di Fisica, Universita' degli Studi di Torino,\\
via Pietro Giuria 1, I-10125, Torino, Italy}
\affiliation{Istituto Nazionale di Fisica Nucleare, Sezione di Torino,\\
via Pietro Giuria 1, I-10125, Torino, Italy}

\author{W. P. Maksym}
\affiliation{Harvard-Smithsonian Center for Astrophysics, \\
60 Garden St., Cambridge, MA 02138, USA}
\affiliation{NASA Marshall Space Flight Center, \\
Martin Rd SW, Huntsville, AL 35808, USA}

\author{M. Karovska}
\affiliation{Harvard-Smithsonian Center for Astrophysics, \\
60 Garden St., Cambridge, MA 02138, USA}




\begin{abstract}

We present \textit{Chandra} ACIS-S imaging spectroscopy results of the extended (1.5$''$-8$''$, 300 pc-1600 pc) hard X-ray emission of NGC 5728, the host galaxy of a Compton thick active galactic nucleus (CT AGN). We find spectrally and spatially-resolved features in the Fe \ka complex (5.0-7.5 keV), redward and blueward of the neutral Fe line at 6.4 keV in the extended narrow line region bicone. A simple phenomenological fit of a power law plus Gaussians gives a significance of 5.4$\sigma$ and 3.7$\sigma$ for the red and blue wings, respectively. Fits to a suite of physically consistent models confirm a significance $\geq$3$\sigma$ for the red wing. The significance of the blue wing may be diminished by the presence of rest frame highly ionized Fe~XXV and Fe~XXVI lines (1.4$\sigma$-3.7$\sigma$ range). A detailed investigation of the \textit{Chandra} ACIS-S point spread function (PSF) and comparison with the observed morphology demonstrates that these red and blue wings are radially extended ($\sim$5$''$, $\sim$1 kpc) along the optical bicone axis. If the wings emission is due solely to redshifted and blueshifted high-velocity neutral Fe K$\alpha$ then the implied line-of-sight velocities are +/- $\sim$0.1$c$, and their fluxes are consistent with being equal. A symmetric high-velocity outflow is then a viable explanation. This outflow has deprojected velocities $\sim$100 times larger than the outflows detected in optical spectroscopic studies, potentially dominating the kinetic feedback power.

\end{abstract}



\section{Introduction}
\label{sec:introduction}
\subsection{The Hard X-ray Emission in AGNs}
\label{sec:origin}
In the standard model of active galactic nuclei (AGNs) \citep{lawrence1982a, antonucci1985a}, the hard X-ray ($>$3 keV) continuum and the 6.4 keV neutral Fe \ka emission line originate from reflection of AGN photons and ensuing fluorescence \citep[e.g.,][]{kallman2007a} in the $\sim$0.1 parsec torus obscuring the nucleus. \textit{Chandra} and \textit{XMM-Newton} grating observations show that the width of the neutral Fe \ka line in unobscured (Type 1) AGNs is $\sim$a few 1000 km~s$^{-1}$, with large error bars \citep{nandra2006a, gandhi2015a, andonie2022a}, consistent with an origin in fluorescence from the inner edge of the dusty torus, or regions at smaller radii \citep{gandhi2015a, andonie2022a}. This line may vary in response to continuum variability \citep{andonie2022a}. In several type 1 Seyfert galaxies, where the view to the nuclear source is unobscured, very broad Fe \ka lines have also been reported, and are explained with relativistic gravitational broadening \citep{tanaka1995a, fabian2000a}. Relativistically broadened lines would originate from physical regions of a few Schwarzschild radii. All these Fe \ka lines of Type 1 AGNs would appear point-like even when imaged with \textit{Chandra}.

However, not all the hard continuum and Fe \ka emission of AGNs is point-like. High spatial-resolution observations of Type 2 CT AGNs in different spectral bands with \textit{Chandra} ACIS have revealed extended (100 pc to kpc scale) components of the hard ($>$3 keV) continuum and neutral Fe \ka line. In these AGNs, the torus obscures all the regions from which the intense Fe \ka lines of Type 1 AGNs are believed to originate. Extended emission in these energy bands have been detected on 100 parsec scales in the nearby AGN NGC 4945 \citep{marinucci2012b, marinucci2017a}, in the Circinus galaxy \citep{andonie2022b}, and in NGC 5643 \citep{fabbiano2018c}, indicating interaction of the AGN photons with structures external to the nucleus, as also observed in the Milky Way \citep{koyama1996a, churazov2016a}. Even more extended, kiloparsec-scale, hard continuum and Fe \ka components have been revealed by deep \textit{Chandra} observations of several other nearby CT AGNs \citep{fabbiano2017a, fabbiano2018a, jones2020a, jones2021a}, and they may be a common feature in a significant fraction of these objects \citep{ma2020a, ma2021a}. This kpc-scale hard X-ray emission suggests interactions of nuclear photons with molecular clouds in the host galaxies, that lie along the ionization cone (directly confirmed by the comparison of \textit{ALMA} and \textit{Chandra} observations in the case of ESO 428-G014; \citealt{feruglio2020a}). Observationally, the detection of these extended components on a few arcsec or less angular scale is only possible in CT AGNs, where the bright nuclear source is reduced by an order of magnitude or more, so that the \textit{Chandra} ACIS detector does not suffer from pile-up and the surface brightness of the PSF wings does not dominate the image \citep{fabbiano2022b}.

One of the CT AGNs in which extended hard X-ray continuum and Fe \ka components have been reported is NGC 5728 \citep[hereafter Paper 1]{trindadefalcao2023a}. NGC 5728 is a barred spiral galaxy located approximately 41 Mpc away (1$''\sim$200 pc, \citealt{mould2000a}), which harbors a highly absorbed CT AGN \citep[log N$_{H}$=24.3 cm$^{-2}$, ][]{koss2017a}. NGC 5728 features a prominent kiloparsec-scale ionization bicone, which arises from the interaction between AGN photons and the interstellar medium (ISM) in the galaxy disk. This AGN also shows a radio jet aligned with the ionization cone \citep{durre2018a}. Deep $\sim$260 ks, \textit{Chandra} observations of NGC 5728 \citepalias{trindadefalcao2023a} unveiled an elongated, kiloparsec-scale, and line-dominated soft (0.3-3 keV) X-ray emission associated with the ionization bicone. \citetalias{trindadefalcao2023a} also reported the discovery of extended hard continuum ($>$3 keV) and rest-frame Fe \ka emission in this galaxy, spatially aligned with the soft emission, as also observed in other CT AGNs with \textit{Chandra} (see review, \citealt{fabbiano2022b}).  

Analyzing separately the X-ray spectra of the nuclear source (inner 1.5$''$) and the extended bicones (1.5$''$-8$''$, 300 pc- 1.6 kpc), \citetalias{trindadefalcao2023a} reported the detection of significant spectral wings of the 6.4 keV neutral Fe \ka line, leading to the present work. Similar features have been recently reported in the \textit{Chandra} X-ray imaging study \citep{maksym2023a} of the CT Type 2 AGN, Mrk 34, where broad spectral wings were associated with the narrow Fe \ka 6.4 keV fluorescence line, within the inner $\sim$200 pc. Imaging of Mrk 34's spectral wings and narrow Fe \ka revealed definite spatial displacements between the components, suggesting very fast outflows seen in emission \citep{maksym2023a}, with estimated line-of-sight velocities of $\sim$15,000 km s$^{-1}$. In this paper, we take a closer look at the spectral Fe \ka wings discovered in NGC 5728, report their spectral and spatial properties, and discuss their possible contribution to a fast nuclear outflow.

\subsection{Multiphase outflows in AGNs}
\label{sec:multiphase}

Outflows and winds from AGNs are thought to be the main drivers of feedback between AGN activity and star formation in their host galaxies \citep[e.g.,][]{king2015a, fiore2017a}, in one of the three forms: (1) Highly ionized winds, with ionization parameter $\xi\sim10^{3}-10^{6}$ erg~cm$^{-2}$s$^{-1}$, are often detected through absorption lines of highly ionized gas, usually Fe XXV and Fe XXVI \citep[e.g.,][]{chartas2003a, tombesi2010a, nardini2015a}, with relativistic velocities in the range $v\sim0.03c-0.3c$ \citep{chartas2003a, pounds2003a, reeves2003a, tombesi2010a}, the so-called Ultra Fast Outflows (UFOs) \citep[e.g.,][]{pounds2003a, tombesi2013a}. UFOs are constrained to be compact ($<$0.03 pc, \citealt{tombesi2012a}) or even to arise at accretion disk scales \citep{gallo2011a}. (2) Another class of AGN winds detected through X-ray absorption of highly ionized species are known as Broad Absorption Lines (BALs), usually observed as blueshifted C~IV, O~VI, N~V, and S~VI species, with line widths implying $v\sim$ 0.1$c$. BALs can be found in approximately 10-30\% of quasars \citep{netzer2013a}, at 10s-100s of parsec-scales from the continuum source \citep{arav2018a}. (3) Moderately ionized winds ($\xi\leq100$ erg~cm$^{-2}$s$^{-1}$) can also be detected through X-ray absorption, in this case appearing in the soft X-ray band as blueshifted absorption lines with velocities ranging from $v\sim$100s-1000s km~s$^{-1}$ \citep{halpern1984a, krongold2003a, kaastra2014a}, and are known as Warm Absorbers (WAs). These winds have small column densities ($\tau_{\rm Compton}\sim$ 0.01), and are thought to be located at larger distances from the SMBH.

Current models of a wind shock \citep[e.g.,][]{king2015a} predict that inner UFOs transfer kinetic energy to the host ISM when shocking against the ambient medium, possibly driving efficient feedback into the galaxy. These models predict that in the aftermath of the shock, different regions or phases arise within the outflow. Besides the inner UFO wind, a ``shocked UFO region" is formed, followed by a region containing the swept-up ISM, and by the outer ambient medium, not yet affected by the inner disk winds \citep[][Section 6.4]{king2023a}. Recently, \citet{serafinelli2019a} identified three different types of absorbers coexisting in the quasar PG 1114+445 \citep{serafinelli2019a}, suggesting the existence of a multi-phase and multi-scale outflow in this source. The first component is a high-velocity ($v\sim0.145c$), high ionization (log $\xi\sim 4$ erg~cm$^{-1}$~s$^{-1}$), high column density (log N$_{H}\sim$23 cm$^{-2}$) component, consistent with the inner UFO. At larger distances ($\sim$100pc), the UFO shocks and entrains the ISM, which is accelerated to comparable velocities ($v\sim$0.12$c$), maintaining its high ionization state (log$\xi\sim 0.5$ erg~cm$^{-1}$~s$^{-1}$) but lower column density (log N$_{H}\sim$21.5 cm$^{-2}$), giving rise to what they refer to as ``extended-UFO". Further out, at kpc-scales, the ambient ISM remains unaffected by the wind, and the third absorber is identified as a low velocity ($v\sim 530$km~s$^{-1}$), moderate ionization (log$\xi\sim$0.3 erg~cm$^{-1}$~s$^{-1}$), moderate column density (log N$_{H}\sim$21.9 cm$^{-2}$) component, consistent with WAs. 
\medskip

Observational evidence for UFOs, BALs, and WAs all rely on the detection of absorption features in the X-ray spectra of luminous, unobscured AGNs. Deep high-spatial and high-spectral resolution observations of nearby obscured AGNs provide a complementary view of nuclear winds and of the interaction between AGN photons and jets with the ISM, both gaseous and in molecular clouds. Several papers have been written on this subject, and we refer the reader to \citet{fabbiano2022b} for a comprehensive review of these observational results. As discussed in Section \ref{sec:origin} above, a recent \textit{Chandra} imaging study of Mrk 34 \citep{maksym2023a} suggests that UFOs and/or BALs may also be detected in emission. In this work, we show that the wings discovered in NGC 5728 may provide a better resolved kpc-scale example.

This paper is organized as follows: in Section \ref{sec:summary}, we summarize the main properties of the source NGC 5728, also reported in \citetalias{trindadefalcao2023a}. In Section \ref{sec:wings_definition}, we analyze the extended bicone spectrum of this AGN, fitting it with a set of emission models to establish the existence and characteristics of the spectral wings. We also compare the spectrum against reflection models to probe the contribution of a possible Compton shoulder emission to the red wing. In Section \ref{sec:images}, we discuss the spatial analysis of the red and blue features, revisiting the comparison of the data with the \textit{Chandra} ACIS-S point spread function (PSF). In Section \ref{sec:disc_conc}, we discuss possible interpretations of our results, and their implication to the overall picture of AGN feedback in the local Universe. In Section \ref{sec:conclusions}, we summarize the main results and conclusions of the results presented in this paper. Appendix \ref{sec:appa} contains the tables of fit parameters for the many models tried in an attempt to fit the spectral wings. Appendices \ref{sec:appb} and \ref{sec:appc} detail the \textit{Chandra} ACIS-S PSF calibration relevant to this paper.

\section{Summary of Previous Results}
\label{sec:summary}
In \citetalias{trindadefalcao2023a}, we presented the results of deep \textit{Chandra} ACIS-S observations of NGC 5728, showing that the X-ray emission from the AGN is extended both in the bicone direction (the direction of the ionization cones) and in the cross-cone direction (the perpendicular direction). The X-ray emission in the bicone direction is extended in the full 0.3-7.0 keV energy range, whereas at higher energies ($>4$ keV) the cross-cone emission matches that of the \textit{Chandra} ACIS-S Point Spread Function (PSF). 

The nuclear ($r<$1.5$''$) spectrum of the AGN is best fit with a low-photoionization gas phase mixed with a more ionized component. Instead, the extended bicone (1.5$''$<$r$<8$''$) and cross-cone (2.5$''$<$r$<8$''$) spectra are dominated by a mix of photoionization and thermal gas emission. We modeled the extended bicone soft continuum with a simple power-law and the neutral Fe line at 6.4 keV with a Gaussian, and found that the individual NW and SE cone spectra are best fit as a mix of photoionized and thermal gas \citepalias{trindadefalcao2023a}.

\citetalias{trindadefalcao2023a} also reports the detection of two unidentified features in both individual cone spectra, to the red and to the blue of the neutral Fe \ka line at 6.4 keV, which we will refer to as red and blue wing, respectively. Physically motivated photoionization and thermal models partially fit the emission in the blue wing as highly ionized iron lines, such as Fe XXV and Fe XXVI, but with significant residual excesses. The red wing remains unmodeled by any combination of photoionized and/or thermal gas models, appearing as residual in all fits. In this paper, we explore the robustness of these results to the use of different spectral models (Section \ref{sec:wings_definition}) and investigate the spatial properties of these spatially and spectrally-resolved features (Section \ref{sec:images}). \par

\section{\element{Fe} \ka Wings in the Bicone Spectrum}
\label{sec:wings_definition}

\begin{figure}
    \centering
    \includegraphics[width=10cm]{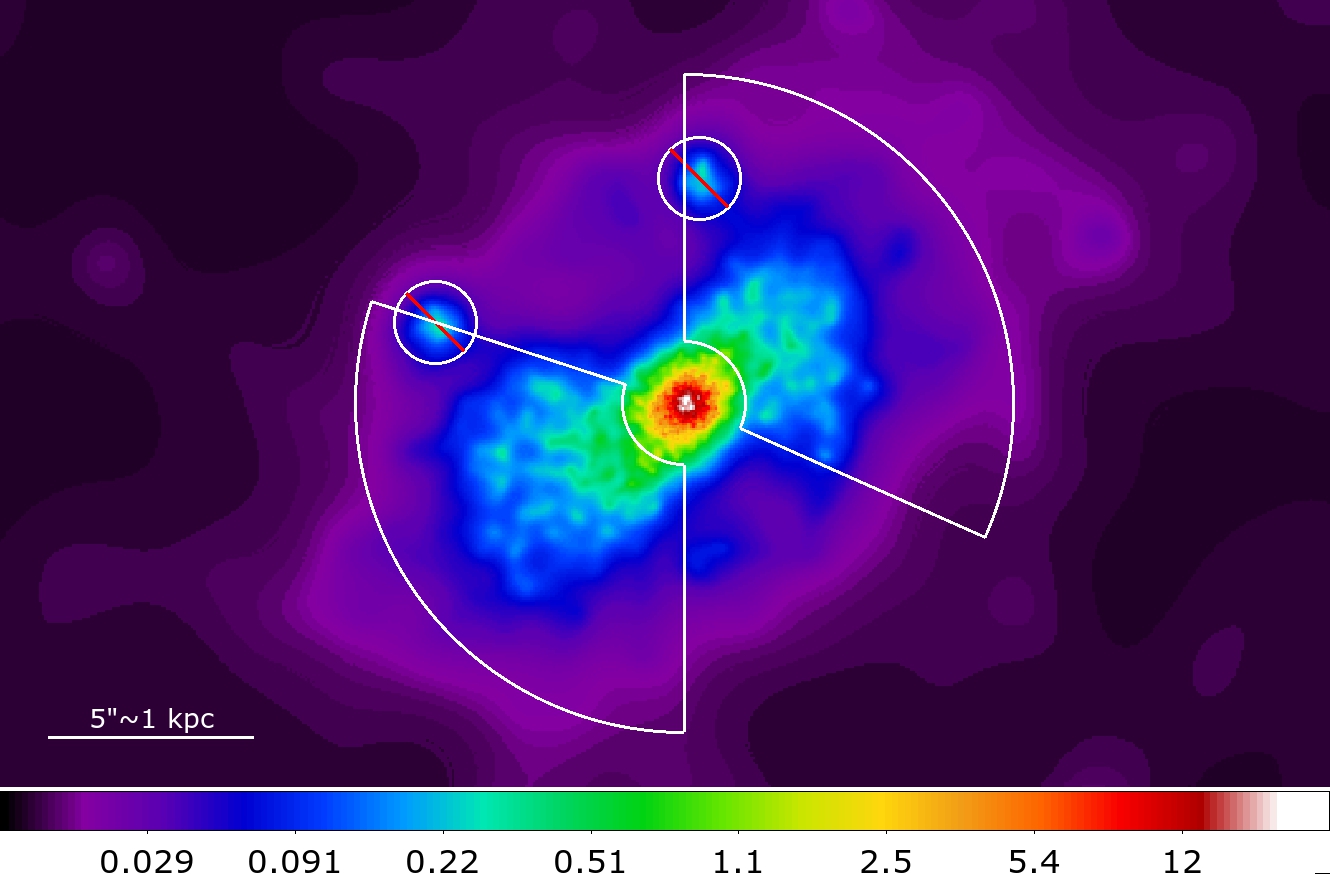}
    \caption{\textit{Chandra} ACIS-S image of the 0.3-8 keV X-ray emission in NGC 5728, at 1/8 pixel. The conical regions are the extraction regions used in the spectral fitting analysis, with an inner, outer radius of 1.5$''$, 8$''$, corresponding to $\sim$300pc, 1,600pc at this redshift. The off-nuclear point sources removed from the spatial and spectral analysis are also shown. The color scale is in units of counts per unit pixel. The image is adaptively smoothed.}
    \label{fig:ds9_regions}
\end{figure}

To improve the statistical robustness of the data, we combined the two individual NW and SE cone spectra, henceforth referred to as the bicone spectrum. We used \texttt{CIAO} \textit{specextract} to extract the spectra from individual observations, and individual spectra and responses were co-added using the \textit{combine\_spectra} script. We then use \texttt{Sherpa}\footnote{https://cxc.cfa.harvard.edu/sherpa/} \citep{freeman2001a}  to fit the final merged bicone spectrum. Fig. \ref{fig:ds9_regions} shows the spectra extraction regions superimposed on the 0.3–8 keV image of NGC 5728. We use the cone regions as defined in \citetalias{trindadefalcao2023a}, with inner and outer radii of 1.5$''$ and 8$''$ (300 and 1,600 pc), respectively, and exclude the regions containing the two bright point sources detected in \citetalias{trindadefalcao2023a} (Fig. \ref{fig:ds9_regions}). The background was extracted from an off-source, circular region of 10$''$ radius, free of X-ray point-like sources. The bicone extraction region yields 2,625$\pm$51 net (background-subtracted) counts in the 0.3-8 keV energy band, which we use for the spectral analysis. 

At energies $>$ 3 keV, the spectrum of the extended bicone emission appears as a smooth continuum with a set of Gaussians superimposed (Fig. \ref{fig:pheno}, in blue). The significance of these bumps, and especially the red and blue wings to the Fe \ka, depends primarily on the correct placement of the continuum. Hence, it is important to try different plausible models to see if any of them systematically erodes the significance of the wings. In Section \ref{sec:pheno}, we fit the extended bicone spectrum with the simplest phenomenological model, a \underline{power-law plus Gaussians}. In Section \ref{sec:physicalmodels}, we explore a mix of models motivated by plausible \underline{physical mechanisms}. These models typically link continuum and emission lines, and the main constraint of these fits is the reproducibility of the soft $<$3 keV emission line spectrum, which will affect the hard continuum. In Section \ref{sec:CS}, we fit the bicone spectrum with a \underline{set of reflection models} to probe the contribution of Compton shoulder emission on the emission in the red wing. These models and spectral analysis are described below. The results demonstrate the robustness of the Fe \ka wings (especially the red wing) to a wide range of plausible emission scenarios.


\subsection{Phenomenological Models}
\label{sec:pheno}

We initially employed three classes of phenomenological models to pinpoint regions of line emission within the extended bicone spectrum. These models include a power-law continuum with varying photon index and several redshifted Gaussian lines. The Gaussian line widths are left free to vary but are restricted to having widths $\geq$ 0.1 keV, corresponding to the ACIS-S spectral resolution.\\

\begin{figure}[htb]
  \centering
 \includegraphics[width=18cm]{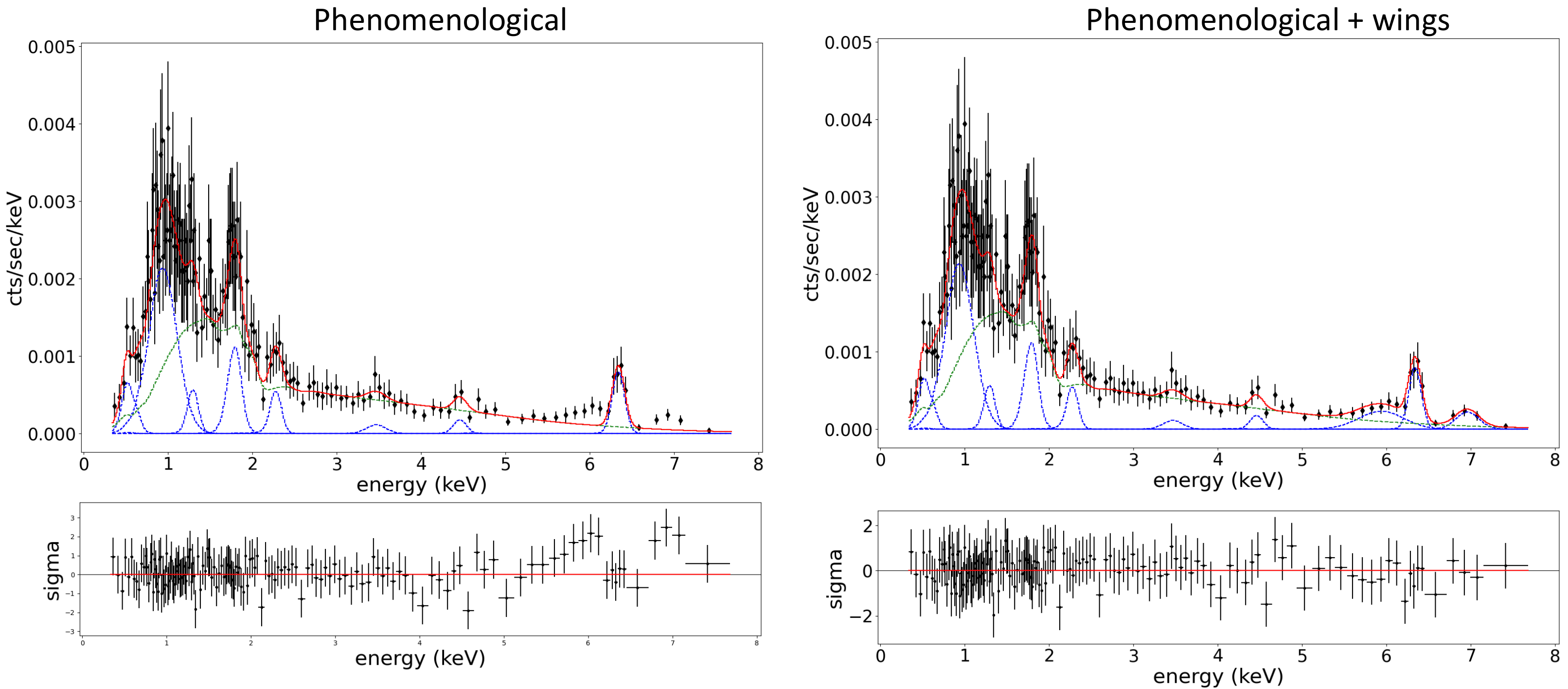}
\caption{\textit{Chandra} ACIS-S 0.3-8.0 keV bicone spectrum of NGC 5728. \textbf{Left:} The continuum is best fit with a power-law with $\Gamma_{\rm pheno\_1}$~=~1.3~$\pm$~0.6. (green line). We modeled the neutral Fe \ka transition along with other prospect emission lines using simple Gaussians (in blue). Note the excesses redward (5.0-6.3 keV) and blueward (6.5-7.5 keV) of the neutral Fe \ka line. \textbf{Right}: We include two Gaussian lines to model the red and blue wing features to the neutral Fe \ka line. In this case, the continuum is best fit with a power-law with $\Gamma_{\rm pheno\_2}$~=~1.4~$\pm$~0.4 (green line). The addition of the red and blue Gaussian components significantly reduced the excesses seen previously in the Fe \ka band.}
 \label{fig:pheno} 
 \end{figure}

\begin{enumerate}
  \item We started by fitting eight Gaussians to the bicone spectrum, as summarized in Table \ref{tab:pheno_models}. Note that this first phenomenological fit does not include Gaussians to model the red and blue wings. The power-law component yielded a photon index $\Gamma_{\rm pheno\_1}$~=~1.3~$\pm$~0.6. The phenomenological power-law components employed in this Section to model the continuum should not be considered ``physical", as they account for a combination of the reflection power-law + soft continuum power-law components. The eight lines fitted here to the bicone spectrum were also detected in the phenomenological models described in \citetalias{trindadefalcao2023a}, where the two sides of the bicone were fitted separately.\par
  The left panel of Fig. \ref{fig:pheno} shows residual excesses at the energies corresponding to the red (5.0-6.3 keV, significance of 5.4$\sigma$) and blue (6.5-7.5 keV, significance of 3.7$\sigma$) wings. 

  \item We then added two Gaussians to model the red and blue wing features (Table \ref{tab:pheno_models}), resulting in a reduction of residuals, as shown in Fig. \ref{fig:pheno} (right panel). The Gaussians have energies, $E_{\rm red}$=6.0$\pm$0.3 keV, and $E_{\rm blue}$=7.0$\pm$0.2 keV, with equivalent widths $EW_{\rm red}$=1.8$\pm$0.2 keV, $EW_{\rm blue}$=2.6$\pm$0.4 keV. The fit yielded a power-law component with a photon index of $\Gamma_{\rm pheno\_2}$~=~1.4~$\pm$~0.4, and a model flux of $\sim$9.5$\times$10$^{-14}$ erg~s$^{-1}$ in the 0.3-8 keV band. A likelihood ratio test\footnote{https://cxc.cfa.harvard.edu/sherpa/ahelp/plot\_pvalue.html} based on a simulation with 100,000 iterations yields a probability p$<$10$^{-5}$ that a simple power-law continuum plus a narrow Fe \ka line is a less suitable representation of the 0.3-8 keV data than a model including the red and blue Gaussian components. 
  \item We used \textit{\texttt{xspexmon}}\footnote{``xs" denotes that the model was used within \texttt{Sherpa}; the model code is identical to the version in \texttt{xspec}.} \citep{nandra2007a}, a neutral reflection spectral model (see Sections \ref{sec:physicalmodels} and \ref{sec:CS}) to fit the underlying power-law continuum, and the neutral fluorescence Fe lines, self-consistently. In this case, we included in the model all the phenomenological Gaussians listed in Table \ref{tab:pheno_models} with energies $<$5 keV. The model yielded a power-law component $\Gamma_{\rm pheno\_3}$~=~1.2~$\pm$~0.8. One emission line was fitted at 6.4 keV, and one line at $E$=6.9 keV, with the latter leaving 3.1$\sigma$ residuals in the blue wing band. No emission lines were fitted to the red wing, leaving a 5.3$\sigma$ residual. 

\end{enumerate}

\subsection{Physically Motivated Models}
\label{sec:physicalmodels}
\begin{figure}
    \centering
    \includegraphics[width=11.5cm]{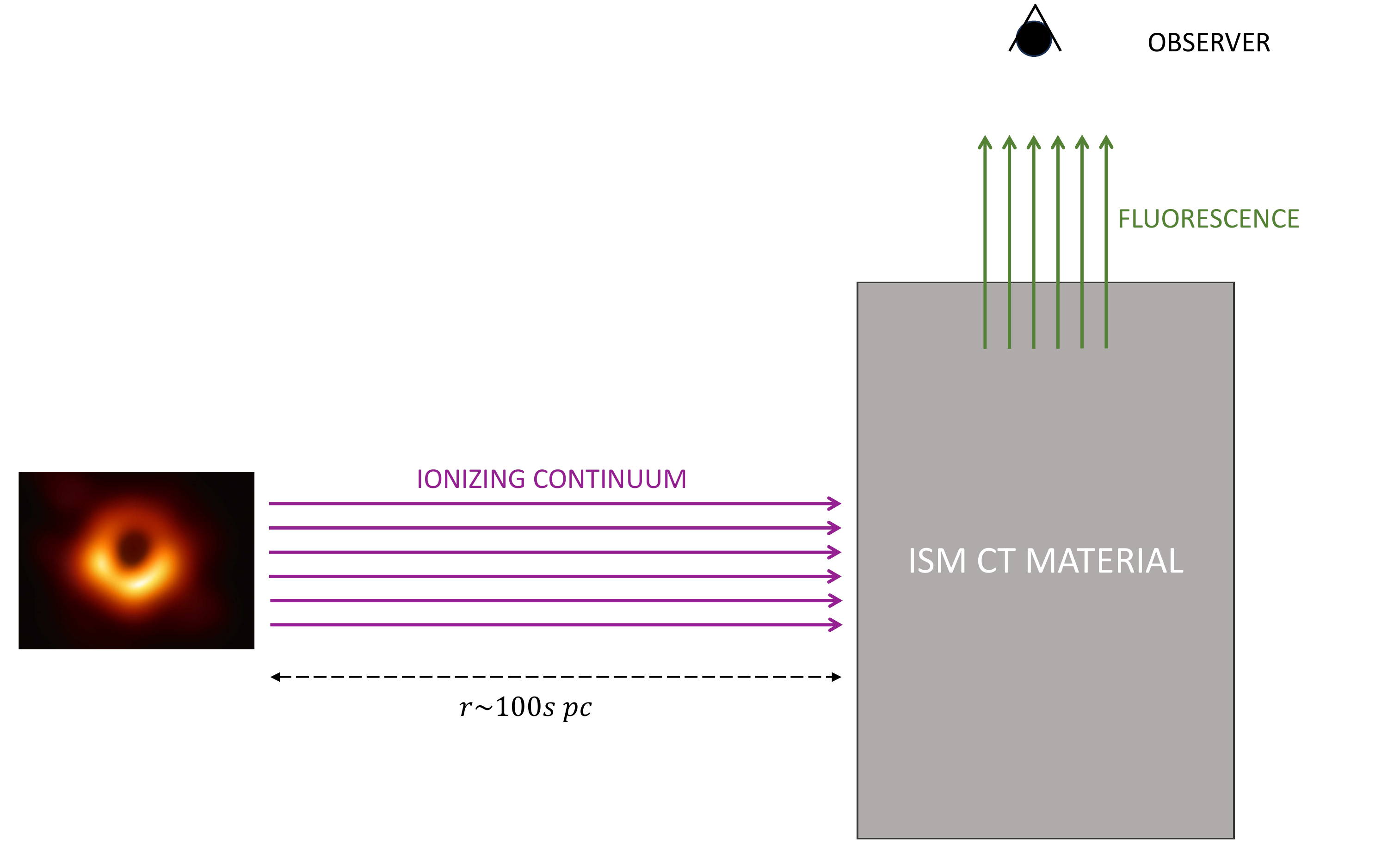}
    \caption{One of the proposed geometry for the extra-nuclear bicone fluorescent emission discovered in NGC 5728. In this scenario, the reflector is material located several 100s of pc outside of the nuclear region of the AGN, as represented by the gray slab in the figure. The CT material gives rise to the observed redshifted/blueshifted fluorescence lines (shown in green) when illuminated from one side by the continuum source (ionizing continuum, in purple).}
    \label{fig:geometry}
\end{figure}

Section \ref{sec:pheno} (and \citetalias{trindadefalcao2023a}) show that several emission lines are present in the spectrum. The mechanisms that could produce these lines in the extended bicones are photoionization and thermal emission. To probe these emission mechanisms, we employed combinations of \texttt{CLOUDY} \citep{ferland2017a} and \texttt{APEC} \citep{foster2012a} spectral fitting models, to model \textit{photoionized and thermal (possibly shocked) gas}, respectively (see also \citetalias{trindadefalcao2023a}). As previously suggested by other studies of nearby CT AGN with \textit{Chandra} \citep[e.g.,][]{fabbiano2017a, paggi2022a}, we assumed a scenario of reflection off molecular clouds in the plane of the galaxy (Fig. \ref{fig:geometry}), and included the reflection model \textit{\texttt{xspexmon}} \citep{nandra2007a}.  \textit{\texttt{xspexmon}} combines the output power-law continuum with self-consistently generated emission lines, such as Fe K$\alpha$, Fe K$\beta$, Ni K$\alpha$, and the Fe K$\alpha$ Compton shoulder (for a more complete discussion of the Compton shoulder analysis see Section \ref{sec:CS}). This reflection model assumes a slab geometry, fitting an exponentially cut-off power law spectrum reflected from neutral material. 

The modeling of the nuclear continuum reflection in CT AGNs commonly employs the use of geometric torus models, such as \texttt{MYTorus} \citep{murphy2009a} and \texttt{borus02} \citep{balokovic2018a}, which assume a reprocessing medium shaped as a torus (or donut) surrounding the continuum source, with variable covering factor. Such description of the reflector is appropriate for modeling the nuclear emission, as done in \citetalias{trindadefalcao2023a} for the inner 1.5$''$ spectrum, but \underline{should not} be employed to probe the properties of the extended emission, since the extraction regions \textit{explicitly} exclude emission from the inner $1.5''$. However, for completeness, in Section \ref{sec:CS}, we show the results of spectral fits applying such models.

\subsubsection{\texttt{(CLOUDY+APEC)+\textit{xspexmon+softpowerlaw}}}
\label{sec:pwrlw_models}

In this Section, we fit the 0.3-8 keV extended bicone spectrum with a combination of photoionized and shocked gas models, as done in \citetalias{trindadefalcao2023a}. We included a power-law to model the soft continuum (\textit{softpowerlaw}), and added \textit{\texttt{xspexmon}} to fit the reflection continuum at higher energies and the neutral Fe \ka line at 6.4 keV. A component accounting for absorption by the Galactic column density along the line of sight (N$_{H}$ = 7.53$\times$10$^{20}$ cm$^{-2}$, derived with the \texttt{nasa heasarc} tool) was included in all models. 

Using \textit{\texttt{xspexmon}} to model solely the reflection power-law component (setting ${\rm rel_{refl}}<0$), we assume a high-energy cutoff $E_{\rm cut}$=200 keV, solar abundances, and $\theta_{\rm inc}$=45$\degree$ \citep{semena2019a}. To improve the continuum fit in the 3-5 keV range, we manually added two Gaussian lines, pinpointed by our phenomenological models in Section \ref{sec:pheno} (Table \ref{tab:pheno_models}) at $E1$=3.5 keV (3.4$\sigma$), and $E2$=4.5 keV (3.8$\sigma$), \textit{\texttt{2gauss}}. 

The \texttt{CLOUDY} models built to model photoionization in this source assumed a continuum source with a spectral energy distribution (SED) in the form of a power law L$_{\nu}$ $\propto$ $\nu^{-(\Gamma-1)}$, with $\Gamma$ = 2.0, for 1$\times$10$^{-4}$ eV $<$ h$\nu$ $<$ 13.6 eV, $\Gamma$ = 2.3, for 13.6 eV $<$ h$\nu$ $<$ 500 eV, and $\Gamma$ = 1.5, for 500 eV $<$ h$\nu$ $<$ 30 keV, with exponential cutoffs above and below the limits \citep[e.g.,][]{kraemer2020a}. The element abundances were considered to be 1.4x solar, i.e., (in log, relative to hydrogen, by number): He=-1.00, C=-3.47, N=-3.92, O=-3.17, Ne=-3.96, Na=-5.69, Mg=-4.48, Al=-5.53, Si=-4.51, S=-4.82, Ar=-5.40, Ca=-5.64, Fe=-4.4, and Ni=-5.75 \citep[e.g.][]{trindadefalcao2021a}. The models were built over a range in log U = [-2.00 : 3.00] in steps of 0.1, and column density log N$_{H_{\rm slab}}$ = [20.0 : 23.5] in steps of 0.1, assuming turbulence velocity $v$=100 km~s$^{-1}$ \citep[e.g.,][]{armentrout2007a, kraemer2020a}. The output parameters from each iteration were converted onto additive emission components (ATABLES) in a FITS format \citep{porter2006a}, and we used \texttt{Sherpa} to interpolate between the values in the grid during the fitting process.\par

We used \texttt{APEC} to model the shocked/thermal emission, assuming solar abundances. We fit the extended spectrum with multi-component models, employing an increasing number of photoionized components (from 1 to 3) and 1 thermal component, and an increasing number of thermal components (from 1 to 3) and 1 photoionized component. Models with 2 photoionized + 2 thermal components were also employed.  \par 

\begin{figure}[h!]
  \centering
 \includegraphics[width=18cm]{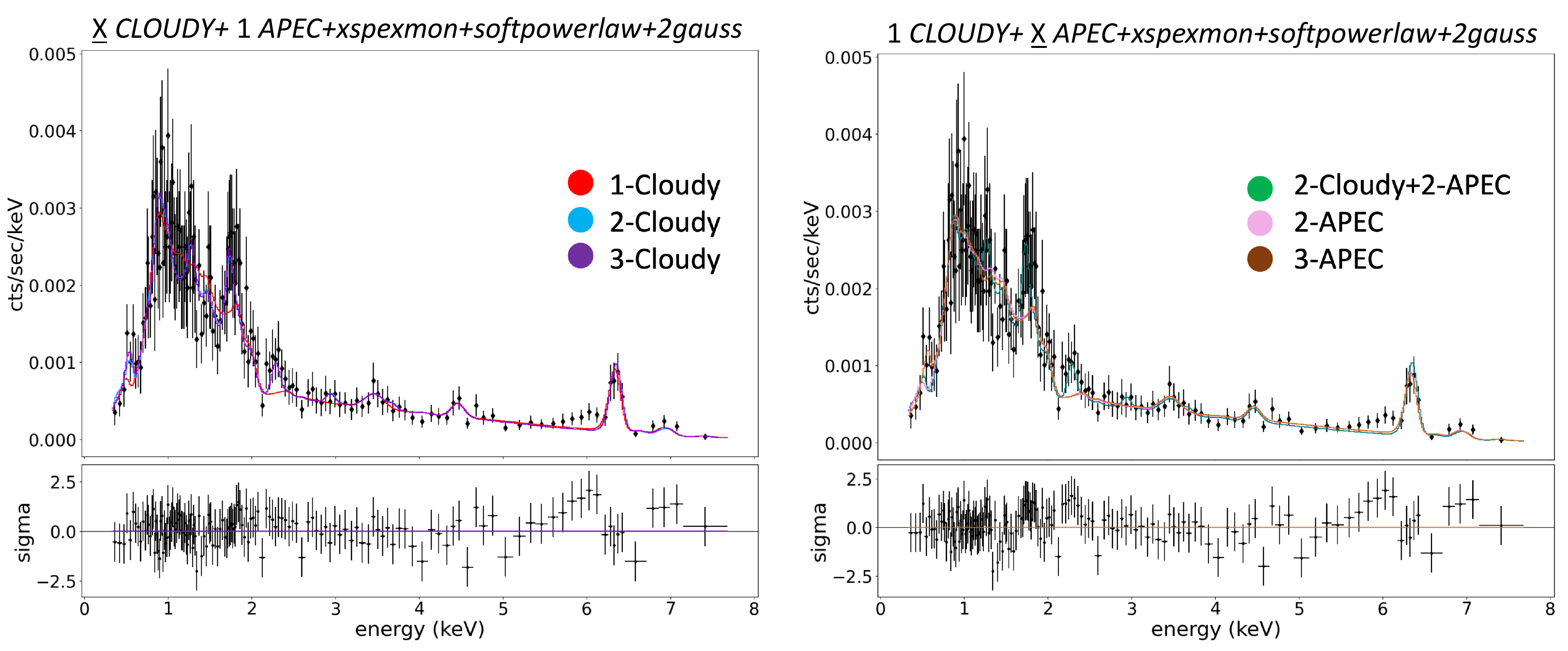}
\caption{\textit{Chandra} ACIS-S 0.3-8.0 keV extended bicone spectrum of NGC 5728. \textbf{Left:} Best-fit models employing \textit{\texttt{xspexmon}}, a soft power-law continuum, 2 Gaussians at $E$=3.5 keV and $E$=4.5 keV, (1,2,3) \texttt{CLOUDY} components + 1 \texttt{APEC} component. Residuals for the (3+1) model fit are shown in the bottom panel. \textbf{Right:} As in the left panel, but models include 1 \texttt{CLOUDY} component + (1,2,3) \texttt{APEC} components, and also a (2+2) combination. Residuals are shown on the bottom panel for the (1+3) model fit.}
 \label{fig:p1} 
 \end{figure}

Fig. \ref{fig:p1} and Table \ref{tab:models} display the results of the models discussed in this Section. Note that the N$_{H_{\rm slab}}$ output from \texttt{CLOUDY} is defined as the column density through the CT material reprocessing the nuclear spectrum, not the line-of-sight column density which is fixed to the Galactic absorption value (N$_{H}$ = 7.53$\times$10$^{20}$ cm$^{-2}$, derived with the \texttt{nasa heasarc} tool). Given the statistics of the data, the reflection photon index ($\Gamma_{\rm ref}$) and the soft power-law photon index ($\Gamma_{\rm soft}$) are unconstrained in all fits.

As discussed in previous studies of CT AGNs with \textit{Chandra} \citep[e.g.,][]{fabbiano2018a}, a simple $\chi^{2}_{\nu}$ criterion should not be used \textit{alone} to determine the goodness of fits in the complex low-energy X-ray spectra. To determine whether additional spectral components were required, we also considered correlated fit residuals that dominated a certain band in the spectrum. This is illustrated in Fig. \ref{fig:p1}, where models with a single photoionized component (red line in the left panel, pink and brown lines in the right panel) fail to fit emission detected at $\sim$1.8 keV and 2.3 keV, and best-fit models require at least 2-photoionization and 1-thermal components to fit the data successfully, all with similar statistics (Table \ref{tab:models} and Fig. \ref{fig:p1}). These multi-component models fit two emission lines to the blue wing, at $E\sim$6.7 keV and $E\sim$6.9 keV, consistent with blended Fe~XXV+Fe~XXVI+Fe~K$\beta$ lines, but with significant residuals ($\geq$2.1$\sigma$). No emission lines were fit to the red wing, which remained evident as residuals in all fits ($\geq$3.2$\sigma$).


\subsubsection{\texttt{(CLOUDY+APEC)+\textit{xspexmon}}}
\label{sec:pex_models}
\begin{figure}[h!]
  \centering
 \includegraphics[width=18cm]{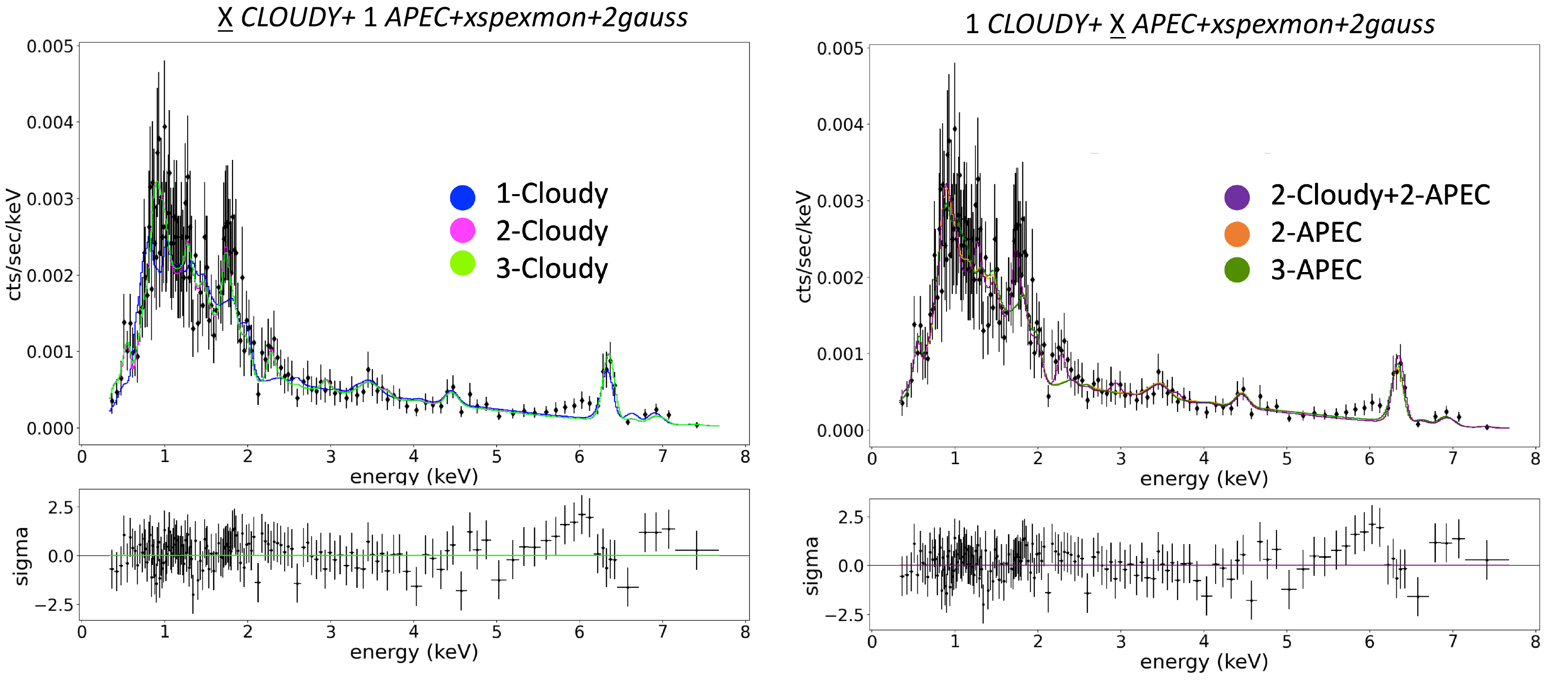}
\caption{\textit{Chandra} ACIS-S 0.3-8.0 keV bicone spectrum of NGC 5728. \textbf{Left:} Best-fit models using \textit{\texttt{xspexmon}}, 2 Gaussians at $E$=3.5 keV, and $E$=4.5 keV, and (1,2,3)\texttt{CLOUDY}+ 1 \texttt{APEC} components. Residuals for the (3+1) model are shown in the bottom panel. \textbf{Right:} Best-fit models including 1 \texttt{CLOUDY}+ (1, 2, 3) \texttt{APEC} components. A model with (2+2) components was also considered, with residuals shown on the bottom panel.}
 \label{fig:pexmon} 
 \end{figure}

A photoionized and/or thermal continuum + emission lines can also fit the extended bicone spectrum, without the addition of an "ad hoc" power-law to model the soft continuum. We fit the 0.3-8.0 keV extended spectrum with models including multi-component \texttt{CLOUDY}+\texttt{APEC}, and \textit{\texttt{xspexmon}} to model the reflection at higher energies. $\Gamma_{\rm ref}$ was left free to vary in all cases. We assumed a fixed high-energy cutoff $E_{\rm cut}$=200 keV, $\theta_{\rm inc}$=45$\degree$, and solar abundances \citep{semena2019a}, and included two Gaussians at 3.5 keV, and 4.5 keV to improve the continuum fit. Best-fit models are shown in Fig. \ref{fig:pexmon}, and described in Table \ref{tab:modelsII}. Models with (2+1), (3+1) or (2+2) \texttt{CLOUDY}+\texttt{APEC} components all provide good fits to the extended bicone spectrum. Two emission lines consistent with blended Fe XXV, Fe XXVI, and Fe K$\beta$ lines were fit to the blue wing, but excesses ($\geq$1.8$\sigma$) remain. The red wing remained unmodeled by any combination of multi-component models tested, leaving $\geq$3.3$\sigma$ excesses, with almost half the cases being $\geq$ 3.9$\sigma$. A model with (1+3) \texttt{CLOUDY}+\texttt{APEC} components fits the blue wing with the smallest residuals (1.4$\sigma$) but represents a worse fit to the soft emission. Given the statistics of our data, $\Gamma_{\rm ref}$ is unconstrained in all fits.

\subsubsection{(\texttt{CLOUDY}+\texttt{APEC)}}
\label{sec:CLOUDY+apec}

\begin{figure}[h!]
    \centering
    \includegraphics[width=18cm]{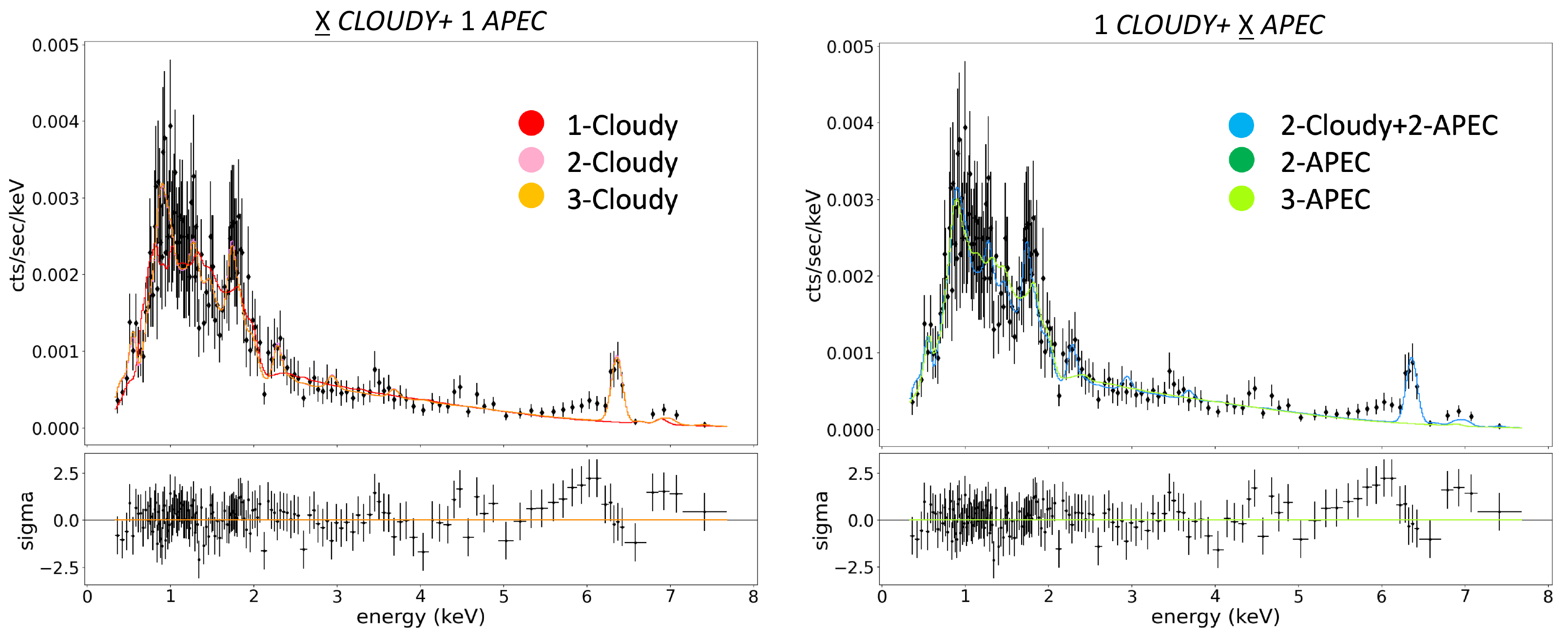}
     \caption{\textit{Chandra} ACIS-S 0.3-8 keV extended bicone spectrum of NGC 5728. \textbf{Left:} Best-fit models using (1,2,3) \texttt{CLOUDY} components+ 1 \texttt{APEC} component, \textit{\texttt{xspexmon}}, and 2 Gaussians at $E$=3.5 keV, and $E$=4.5 keV. Residuals for the (3+1) multi-component model are shown in the bottom panel. \textbf{Right:} Best-fit models using 1 \texttt{CLOUDY} component+ (1, 2, 3) \texttt{APEC} components. A (2+2) model was also considered, with residuals shown on the bottom panel.}
    \label{fig:alternative_models}
\end{figure}

Given that \texttt{CLOUDY} also models fluorescence processes in photoionized gas, in this Section we fit the 0.3-8 keV extended bicone spectrum with multi-component \texttt{CLOUDY}+\texttt{APEC} models. In this case, \texttt{CLOUDY} models the reflection at higher energies. Results are shown in Fig. \ref{fig:alternative_models} and Table \ref{tab:red_wing_alternative}. In all cases, at least 2 photoionized components are required to fit the data. Given the statistics of our data, the soft power-law and reflection power-law photon indexes are unconstrained in all fits.

Models with (3+1), (2+1) and (2+2) \texttt{CLOUDY}+\texttt{APEC} components return the best fits but leave $\geq$4.4$\sigma$ residuals in the red wing, and $\geq$2.4$\sigma$ in the blue wing. The addition of two Gaussians to the model at $E$=3.5 keV and $E$=4.5 keV does not improve the goodness of the fit, as shown in Table \ref{tab:red_wing_alternative}. Models including only 1 \texttt{CLOUDY} component (Fig. \ref{fig:alternative_models}, red and green lines, yielding a high-ionization/low column density component, as shown in Table \ref{tab:red_wing_alternative}) fail to fit the neutral Fe \ka emission at 6.4 keV, and soft emission observed at $\sim$2.3 keV. Models with 2 or 3 \texttt{CLOUDY} photoionization components (Fig. \ref{fig:alternative_models}, pink, orange, and blue lines), return an overall better fit by including a low-ionization/high column density component (see Table \ref{tab:red_wing_alternative}). This high column density (N$_{H_{\rm slab}}$) component is required to model the slab of gas producing the observed neutral Fe emission but with large uncertainties in the related output parameters. For models consisting of 3 photoionized components, an intermediate N$_{H_{\rm slab}}$ is fit to the data, but with no improvements to the goodness of the fit (Fig. \ref{fig:alternative_models}).\par 

Note that \texttt{CLOUDY} photoionization models do not include estimates of Compton shoulder emission. This is addressed in Section \ref{sec:CS}.

\subsection{Effects of the Compton Shoulder Emission on the Red Wing}
 \label{sec:CS}
The Compton shoulder is a spectral feature that arises due to Compton down-scattering of high-energy photons in a high column density medium. This phenomenon has been extensively investigated for the AGN molecular torus with a wide range of properties, employing different methods: analytically by \citet{matt2002a} and \citet{yaqoob2010a}, via Monte Carlo simulations by \citet{george1991a,furui2016a}, and observationally for the Galactic Center (a geometry closer to the situation in NGC 5728) by \citet{odaka2011a}. The results from these works show that:
\begin{itemize}

 \item The ratio between the EW of the Compton shoulder to that of the Fe \ka line, using solar abundances, never exceeds 0.2 in the analytical treatments \citep{matt2003a, yaqoob2010a}.
 \item The Monte Carlo simulations show that if a smooth torus model is considered, this ratio appears to never exceed 0.35, for any given inclination angle \citep[][Fig. 10 therein]{furui2016a}. In a more physically realistic clumpy torus, the ratio never exceeds 0.25 and is typically $\leq$0.2 \citep[][Fig. 15 therein]{furui2016a}. Larger EW ratios ($>$0.4) were found only with sub-solar abundances (of order $\sim$0.2 Z/Z$_{\rm sol}$) at large N$_{H}$ $\sim$ 10$^{24}$ - 10$^{25}$ cm$^{2}$ \citep[][Figs. 10, 11 therein]{furui2016a}.
 \item The width of the Compton shoulder investigated by these authors never extends to energies below 6.2 keV (\citealt[][Fig. 2 therein]{odaka2011a}; \citealt[][Figs. 7, 9, 12, 14 therein]{furui2016a}). 
 
\end{itemize}

 The red wing in NGC 5728 has an EW ratio 0.7$\pm$0.1, \textit{significantly in excess of the largest Compton shoulder estimates.} Moreover, the observed extended red wing extends down to energies $\sim$5.0 keV (Fig. \ref{fig:pheno}), \textit{significantly lower than reported for a Compton shoulder}. \par 

A likelihood ratio test based on a simulation with 100,000 iterations yields a probability p$<2\times$10$^{-3}$ that a Compton shoulder, modeled with the largest possible amplitude (corresponding to 20\% of the Fe \ka EW, see \citealt[][Fig. 15 therein]{furui2016a}) is a better representation of the 5.0-6.3 keV data than a model including a red wing broad Gaussian (see Fig. \ref{fig:pheno}, right). 


Although a Compton shoulder origin to the red wing in the extended bicone seems unlikely, in the next Sections we fit the bicone spectrum in the 3-8 keV energy range of interest, using \textit{\texttt{xspexmon}}, \texttt{MYTorus} \citep{yaqoob2010a} and \texttt{borus02} \citep{balokovic2018a} to model the reflection component. The models considered span different angles of incidence, and iron abundances (but see Section \ref{sec:physicalmodels} about how \texttt{MYTorus} and \texttt{borus02} do not strictly apply to the bicone geometry analyzed in this paper).

\subsubsection{\textit{\texttt{xspexmon}} Reflection Models}

\begin{figure}[h!]
    \centering
    \includegraphics[width=18cm]{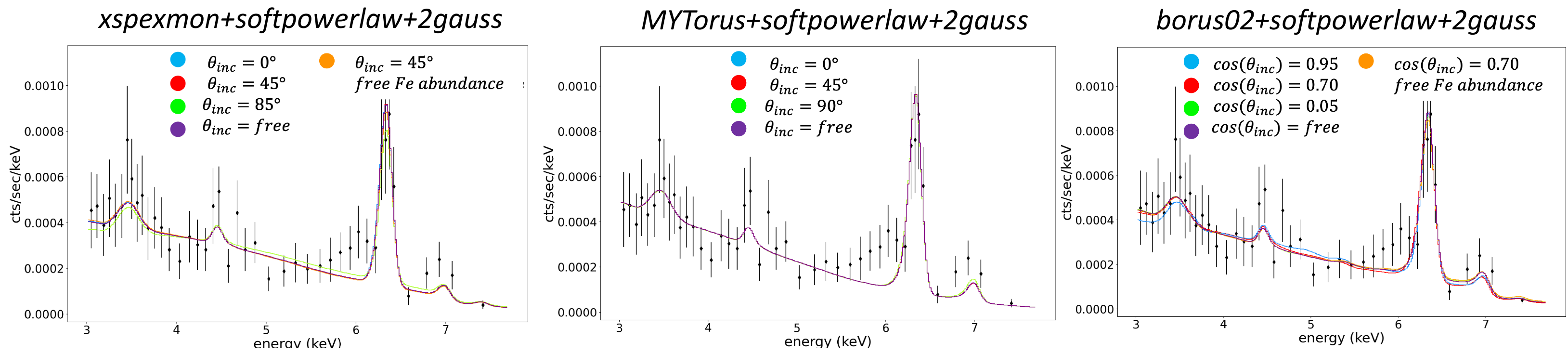}
     \caption{\textit{Chandra} ACIS-S 3-8 keV bicone spectrum of NGC 5728. \textbf{Left:} Best-fit models using \textit{\texttt{xspexmon}} reflection+\textit{\texttt{softpowerlaw}} continuum+\texttt{2gauss}, with varying $\theta_{\rm inc}$. A model with free Fe abundances and $\theta_{\rm inc}$=45$\degree$ was also considered. \textbf{Middle:} Best-fit models using \texttt{MYTorus} reflection+\textit{\texttt{softpowerlaw}} continuum+\texttt{2gauss}, with varying $\theta_{\rm inc}$; \textbf{Right:} Best-fit models using \texttt{borus02} reflection+\textit{\texttt{softpowerlaw}} continuum+\texttt{2gauss}, with varying $\theta_{\rm inc}$.}
    \label{fig:cs}
\end{figure}

We first used \textit{\texttt{xspexmon}} to model the reflection component in the 3-8 keV bicone spectrum. Since we fit solely the energy range $>$3 keV, the addition of a power-law to model the soft continuum is not required. We included two Gaussians at $E$=3.5 keV, $E$=4.5 keV, and fit inclination angles, $\theta_{\rm inc}$ of 0$\degree$, 45$\degree$, 85$\degree$, and also a model with $\theta_{\rm inc}$ free to vary. A model with free iron abundances and $\theta_{\rm inc}$=45$\degree$ was also considered. Fig. \ref{fig:cs} (left panel) and Table \ref{tab:red_wing_models} show the results of our model fits. Modeling the reflection component with \textit{\texttt{xspexmon}} leaves $\geq$ 3.1$\sigma$ excess in the red wing in 4 out of 5 fits. The output soft power-law photon index was unconstrained in all cases. The residual excess in the red wing is 2.9$\sigma$ for a model considering an incidence angle of 85$\degree$, the smallest excess yielded by this set of models. Therefore, \textit{\texttt{xspexmon}} does not fit a Compton shoulder to the emission in the red wing, for any of the models considered. In the blue wing,  \textit{\texttt{xspexmon}} leaves $\geq$2.8$\sigma$ excess residuals. However, the continuum is not constrained by the spectrum $<$3 keV. In Tables \ref{tab:models} and \ref{tab:modelsII}, where the soft spectrum is taken into account, the red wing significance is always $>$3.2$\sigma$. Also, the blue wing in Tables \ref{tab:models} and \ref{tab:modelsII} has generally smaller significance because of the ionized Fe K lines. This consideration applies also to the \texttt{MYTorus} and \texttt{borus02} fits described below.
 
\subsubsection{\texttt{MYTorus} Reflection Models}
The \texttt{MYTorus} spectral fitting model \citep{yaqoob2010a} assumes that the reflector is a torus with homogeneous absorbing material, with a fixed opening angle of 60$\degree$, which translates to a covering factor of 0.50. This model is composed of three distinct components: (1) Line-of-Sight Continuum (\texttt{MYTZ}), which models the X-ray emission from the AGN after it has been absorbed by the torus, along the observer's line of sight; (2) Compton-Scattered Continuum (\texttt{MYTS}), which models X-ray photons that interact with the dusty surrounding torus and scatter into the observer's line of sight, and includes the Compton shoulder; (3) Fluorescent Line Emission (\texttt{MYTL}), which models significant fluorescence lines, such as Fe \ka and Fe K$\beta$. 

We used only the reflection (\texttt{MYTS}) and emission line (\texttt{MYTL}) components, as there is no direct continuum at this off-nuclear location.  The reflection power-law photon index ($\Gamma_{\rm ref}$), and the absorbing column density (the equatorial column density of the torus, N$_{\rm H_{\rm eq}}$) were left free to vary. The models fit inclination angles of 0$\degree$, 45$\degree$, 90$\degree$, and we also considered a model with $\theta_{\rm inc}$ free to vary. The results of this set of models are shown in Fig. \ref{fig:cs} (middle panel) and Table \ref{tab:red_wing_mytorus}. All models employing \texttt{MYTorus} show similar statistics, with $\geq$ 3.3$\sigma$ residuals in the red wing, and $\geq$1.9$\sigma$ in the blue wing.

\subsubsection{\texttt{Borus02} Reflection Models}
\label{sec:borus02}
The \texttt{borus02} \citep{balokovic2018a} torus reflection model generalizes \texttt{MYTorus} by including a variable opening angle, and a variable mean torus column density, N$_{\rm H_{\rm tor}}$. \texttt{Borus02} only models the reflected component (i.e., no direct continuum), which includes both the reflected power-law continuum, fluorescence lines, and the Compton shoulder. 

We fit the extended bicone spectrum with \texttt{borus02} to model the reflection at high energies, using a fixed high-energy cutoff $E_{\rm cut}$=200 keV, and assuming solar abundances. The models fit at inclination angles of cos($\theta_{\rm inc}$)=0.95, 0.7, 0.05 (closely equivalent to $\theta_{\rm inc}$ = 0$\degree$, 45$\degree$, 90$\degree$), and also with cos($\theta_{\rm inc}$) free to vary. In all the models considered, the reflection power-law photon index ($\Gamma_{\rm ref}$), the torus mean column density (N$_{\rm H_{\rm tor}}$), and the torus covering factor (C$_{\rm tor}$=cos($\theta_{\rm tor}$)) are left free to vary. The results are shown in Fig. \ref{fig:cs} (right panel) and Table \ref{tab:red_wing_borus02}. A model considering cos$(\theta_{\rm inc})$=0.7 (or $\theta_{\rm inc}$=45$\degree$) and free iron abundances was also considered (see also in Table \ref{tab:red_wing_borus02}). All best-fit models show similar statistics, with smaller $\chi^{2}_{\nu}$ when compared to the results from \textit{\texttt{xspexmon}} and \textit{\texttt{MYTorus}} (but see below). Residual excesses $\geq$2.9$\sigma$ remained in the red wing, and $\geq$2.4$\sigma$ in the blue wing. 

Although \texttt{\textit{borus02}} has been widely used to model torus reflection in CT AGNs, we note that a recent study by \citet{vandermeulen2023a} found contrasting results between model outputs from their spectral fitting code \texttt{SKIRT}, and results obtained using \texttt{\textit{borus02}} \citep{balokovic2018a}. Their results were consistent with results obtained using \textit{\texttt{xspexmon}} and \texttt{\textit{MYTorus}}.

\subsection{Summary of Spectral Fitting Results}
\label{sec:summary_spec}
Section \ref{sec:pheno} yields high significance for the red and blue wings (5.4$\sigma$, 3.7$\sigma$, respectively, Table \ref{tab:pheno_models}). More complex, multi-component, physically motivated models partially fit the emission in the blue wing as blended rest-frame Fe XXV, Fe XXVI, Fe K$\beta$ emission lines, but with $\geq$1.8$\sigma$ residuals (Table \ref{tab:modelsII}). When tested against 33 different models, the red wing remained significant ($>$3$\sigma$ in the great majority of cases, and 2.9$\sigma$ in 5 of the fits, 4 of which are results of \texttt{borus02} fits, but see Section \ref{sec:borus02} for cautionary note). Reflection models (Section \ref{sec:CS}, \textit{\texttt{xspexmon}}, \texttt{borus02}, and \texttt{MYTorus}) show that the Compton shoulder does not contribute significantly to the red wing. However, the fits in Section \ref{sec:CS} are limited to the 3-8 keV band. Without being constrained by the soft emission spectrum, the resulting continuum in the hard band is higher, minimizing the emission in the red and blue wing energy bands.
\\

\section{Spatial Properties of the \element{Fe} \ka Complex}
\label{sec:images}

Thanks to \textit{Chandra}'s sub-arcsecond resolution, we can characterize the spatial properties of the Fe K$\alpha$ complex in NGC 5728 on sub-kpc scales. This characterization deeply depends on a good understanding of the \textit{Chandra} Point Spread Function (PSF). In this Section, we first discuss our method of spatial analysis employed in \citetalias{trindadefalcao2023a} and in other works \citep[e.g.,][]{fabbiano2018a, paggi2022a}, to study extended X-ray emission in CT AGNs. We also justify the use of the \textit{Chandra} X-ray Center (CXC) tool \texttt{marx}\footnote{\texttt{marx} includes the PSF mirror pre-launch calibration and the response of the ACIS-S detector -https://cxc.cfa.harvard.edu/ciao/threads/marx\_sim/}, for generating a PSF model for these types of data. We then apply this spatial analysis to the spatially-extended wings in NGC 5728.

\subsection{Our Approach to the Spatial Analysis of CT AGNs}
\label{sec:method}

By selection, NGC 5728 is a type 2 CT AGN. This selection ensures that the nuclear source is highly dimmed (and totally obscured at low energies). CT AGNs are typically not affected by pileup in ACIS-S observations given their fairly low count rate, but require deep observations to ensure good signal-to-noise.

\citetalias[][]{trindadefalcao2023a} shows that the X-ray emission from this galaxy is elongated, following the optical ionization cone. In the 4-7 keV energy range, in particular, \citetalias{trindadefalcao2023a} (Fig. 5, therein) shows that the bicone radial profile of the cone surface brightness is significantly more extended than that of the \texttt{marx} ACIS-S PSF models at these energies,  considering a model built with \textit{AspectBlur}=0.2$''$\footnote{https://cxc.cfa.harvard.edu/ciao/why/aspectblur.html}, and normalized to the central 0.5$''$ radius of the peak surface brightness. Instead, the azimuthal emission distribution of the cross-cone regions is consistent with the \texttt{marx} ACIS-S PSF model built with \textit{AspectBlur}=0.2$''$. These azimuthal differences show that (1) the \texttt{marx} ACIS-S model PSF is consistent with the data where the image does not show real extent and, therefore, is a good representation of the point response of the telescope and that (2) if we then use this self-calibrated model in the ionization bicone direction, we detect significant and real emission extent.

\begin{figure}[h!]
    \centering
    \includegraphics[width=15cm]{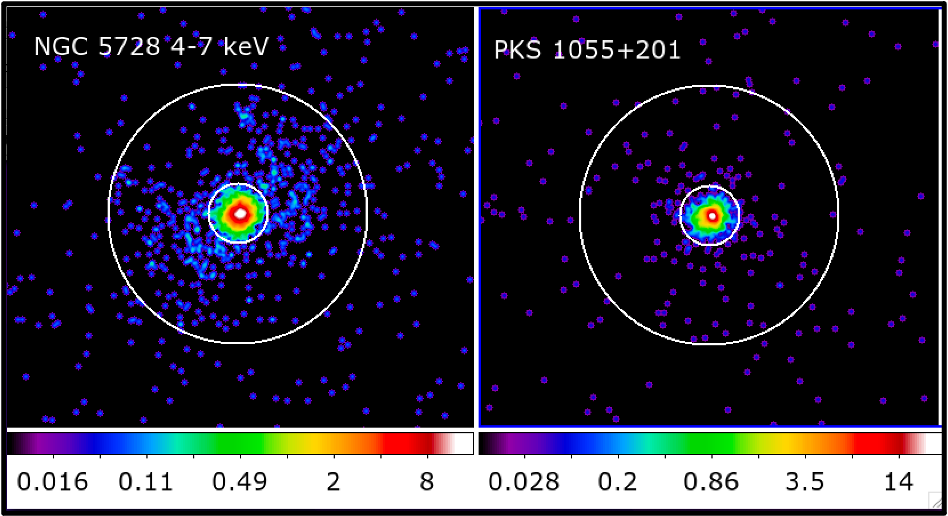}
     \caption{\textit{Chandra} ACIS-S images of the 4-7 keV emission in NGC 5728 (left panel) and PKS 1055+201 (right panerl), on the same scale. Image binning of 1/8 instrumental pixel was used. To increase visibility, the images were smoothed with a 3-image pixel Gaussian. In both panels, the inner circle has $r$=1.5$''$, and the outer circle has $r$=7$''$. The color scale is in counts per image pixel.}
    \label{fig:pks}
\end{figure}

We further note that the extended bicone emission cannot be explained by intrinsic PSF wings, as the \textit{Chandra} ACIS-S PSF at the aim point is azimuthally symmetric (see the \textit{Chandra} Observer Guide\footnote{https://cxc.cfa.harvard.edu/proposer/POG/} and \citet{andonie2022a}, Fig.10 therein). We further probed this point by comparing NGC 5728's data in the 4-7 keV range to an empirical PSF, reaching similar conclusions. For this comparison, we searched the CXC Source Catalog (v.1.1; single observation data) for the best point source empirical PSFs, according to the following criteria: highly variable sources (by a factor $>$ 2 to ensure a dominant point source), within 1 arcminute of the aim point, with a hard-band count rate $<$ 0.1 counts/s to ensure weak pile-up (<10\%), and high number of hard band counts ($>$ 2 keV) in a single ACIS-S observation ($>$ 3,000). Of the resulting list, we chose the source with the highest number of hard-band counts (9,300), which satisfies all criteria described above, and is moreover spatially isolated. This source is a flat spectrum radio source, the quasar PKS 1055+201 (ACIS- ObsID 7795). Fig. \ref{fig:pks} shows the images of NGC 5728 and PKS 1055+201, both in the 4-7 keV band.

Visually comparing these two images, it is clear that the 4-7 keV emission in NGC 5728 cannot be due to a single nuclear point source. The image of PKS 1055+201 shows a centrally peaked, azimuthally symmetric count distribution, quite different from the elongation present in NGC 5728. In both cases, most of the nuclear counts are found within a circle of $r$=1.5$''$, centered on the emission peak: 3,075 for NGC 5728 and 2,578 for PKS 1055+201; in both images, $\sim$4 counts were found in a circle with the same radius, in a source-free region of the field. Using PKS 1055+201 as our empirical PSF, we evaluated the ratio of net counts detected in the 1.5$''$-7$''$ annulus to those within a circle $r$=1.5$''$ (including the core of the PSF). The results showed that 4.5\% $\pm$0.22\% of the core counts were located in this circular region. A similar analysis for NGC 5728 gives a significantly larger percent of counts in the outer annulus, 15.1\% $\pm$ 0.01\%. Normalizing the core counts of PKS 1055+201 to those of NGC 5728, showed that $\sim$138$\pm$12 counts would be expected from the PSF in the outer annulus of NGC 5728, compared against the observed $\sim$463$\pm$ 22 counts, \textit{a highly and significant $\sim$14$\sigma$ difference.}

We further compared the expected counts from the empirical PSF (using PKS 1055+201) in the cone and cross-cone regions, in the 4-7 keV energy band. Using the same cone and cross-cone angles as in \citetalias{trindadefalcao2023a}, we obtained 55 and 59 net counts from the cone and cross-cone regions for PKS 1055+201 between 1.5$''$ and 7$''$, and 383 and 87 net counts from the same regions in NGC 5728. The expected PSF wing contribution (normalized to the total counts within $r$=1.5$''$) are $\sim$67 and $\sim$70 counts respectively. Therefore, the cross-cone emission of NGC 5728 is within 1.4$\sigma$ of the wings of the empirical PSF, in agreement with the analysis of \citepalias{trindadefalcao2023a}, which used the \texttt{marx} PSF. Instead, the bicone emission has 317 counts, or a $\sim$15$\sigma$ excess (Table \ref{tab:counts}).

\begin{table}[h!]
\caption{Comparison between NGC 5728, PKS 1055+201 and the \texttt{marx} model PSF}
\label{tab:counts}
\begin{center}       
\begin{tabular}{|c|c|c|c|c|}
    
    \hline
    & \textbf{Bicone counts} & \textbf{Excess (counts)} &  \textbf{Cross-cone counts}& \textbf{Excess (counts)} \\
    & (4-7 keV) & & (4-7 keV) &  \\ 
    \hline
     
    NGC 5728 & 383&317 (15$\sigma$) & 87&17 (1.4$\sigma$) \\
    \hline

    PKS 1055+201$^{a}$ & 55&& 59& \\
    \hline

     \texttt{marx} PSF wings$^{a}$& 67&& 70& \\
    \hline

\end{tabular}
\end{center}
\centering
\footnotesize$^{a}$ PSF normalized to nuclear region of NGC 5728 ($r<1.5''$)
\end{table} 

In Appendix B, we provide a detailed comparison between the \texttt{marx} model PSF and the empirical PSF (PKS 1055+201) used here, which validates the use of the \texttt{marx} PSF model for our spatial analysis.

\subsection{Spatial Analysis of the Spectral Wings of NGC 5728}
\label{sec:images_subsection}
\begin{figure}[h!]
  \centering
  \includegraphics[width=16cm]{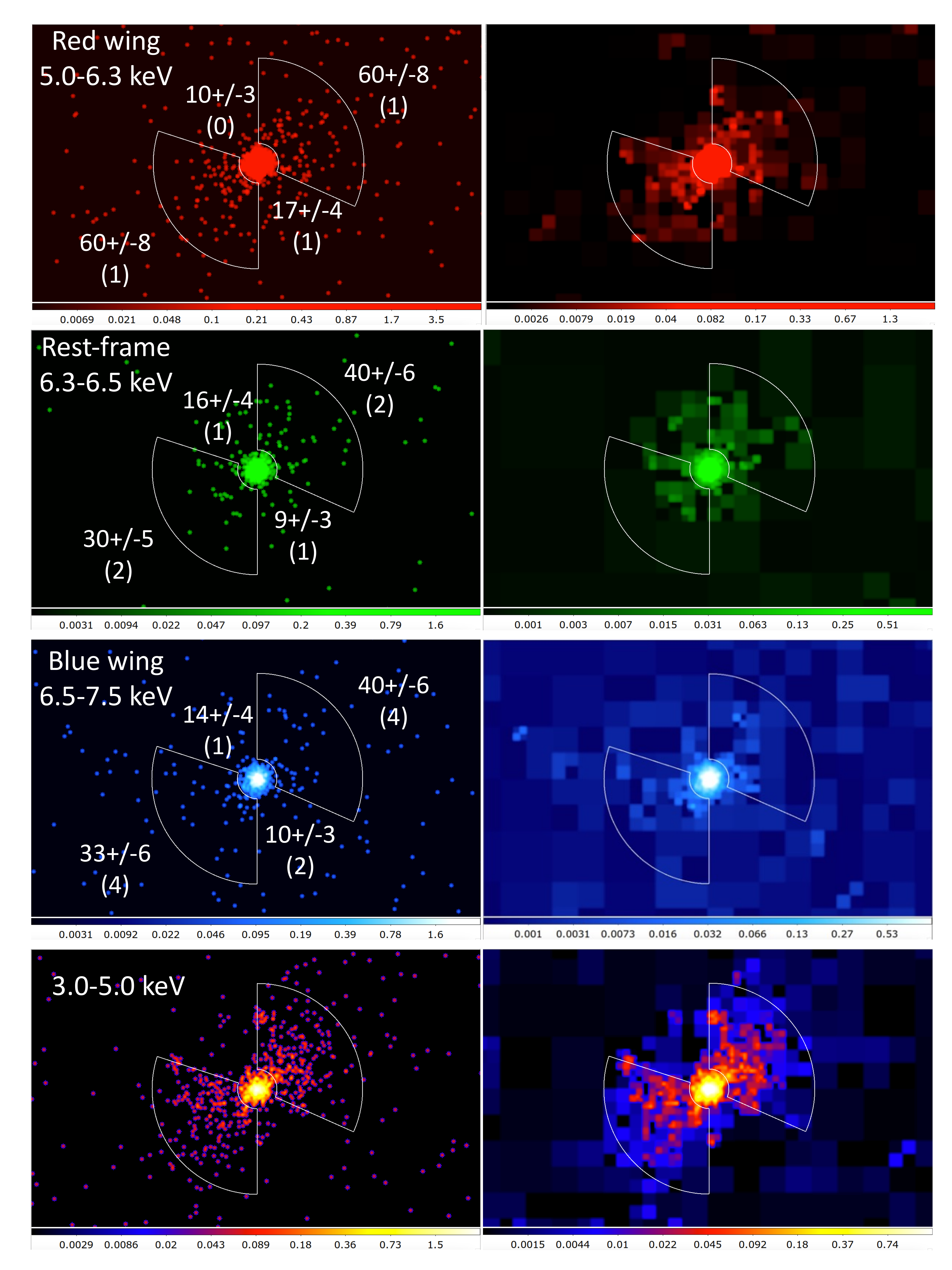}
  \caption{1/8 pixel \textit{Chandra} ACIS-S images of NGC 5728 \textbf{Left:} Red wing (5.0-6.3 keV, top row), rest-frame (6.3-6.5 keV, top-middle row), blue wing (6.5-7.5 keV, bottom-middle row), and 3-5 keV continuum (bottom row) bands. Images are in log intensity scale. The cone sectors, with [1.5$''$, 8$''$] (or 300pc-1,600pc) [inner, outer] radius (including the bulk of the extended emission), are also shown. The cone opening angles are 114$\degree$ for the NW cone and 108$\degree$ for the SE cone. The number of counts in each cone sector (excluding the counts from an inner circle of $r$=1.5$''$) is shown, and in parenthesis are the expected field background counts in these areas. \textbf{Right:} 1/8 pixel adaptively binned images for the shown energy bands (S/N=3).} 
 \label{fig:image} 
\end{figure}

To analyze whether the blue and red wings have spatial extent, we used \texttt{CIAO} \texttt{ds9} to spatially rebin the data with a resolution of 1/8 of the instrumental pixel, a standard technique \citep[see e.g.,][]{fabbiano2022b}, and created images of the extended X-ray emission in the Fe K$\alpha$ rest frame (6.3-6.5 keV), red wing (5.0-6.3 keV), blue wing (6.5-7.5 keV), and 3-5 keV continuum bands. The 3-5 keV continuum band was chosen to image the circumnuclear molecular clouds responsible for reflecting AGN photons that escape along the ionized bicone, excluding any line contribution. The resulting images are shown in Fig. \ref{fig:image}, with the two conical sectors used for spectral extraction, which both exclude the inner $r$=1.5$''$ nuclear emission. For each narrow-band image, Fig. \ref{fig:image} also gives the background subtracted number of counts in each angular sector and, in parentheses, the estimated contribution from the 3-8 keV hard continuum, based on the measured continuum counts and extrapolation to the band of interest, using the spectral fits from \citetalias{trindadefalcao2023a}.

The red wing (5.0-6.3 keV) has $\sim$3$\times$ the counts of the continuum in this energy band and an elongated morphology in the same direction as the 3-5 keV continuum. The rest frame Fe K$\alpha$ (6.3-6.5 keV) and blue wing (6.5-7.5 keV) are similarly extended along the direction of the 3-5 keV continuum and ionization bicone.

Following \textit{Chandra} science threads\footnote{https://cxc.cfa.harvard.edu/ciao/threads/}, we used \textit{dmextract} to extract emission radial profiles for these energy bands and for the wings of the \textit{Chandra} PSF, normalized to the narrow nuclear emission in the band. A 5$''$ radius circular region, free of point sources, was used to extract the background. We used a single power-law ($\Gamma$=1.5) and one Gaussian line at 6.4 keV to estimate the continuum contribution in the 5-8 keV, and normalized the continuum counts by this factor. 

\begin{figure}[h!]
  \centering
 \includegraphics[width=15cm]{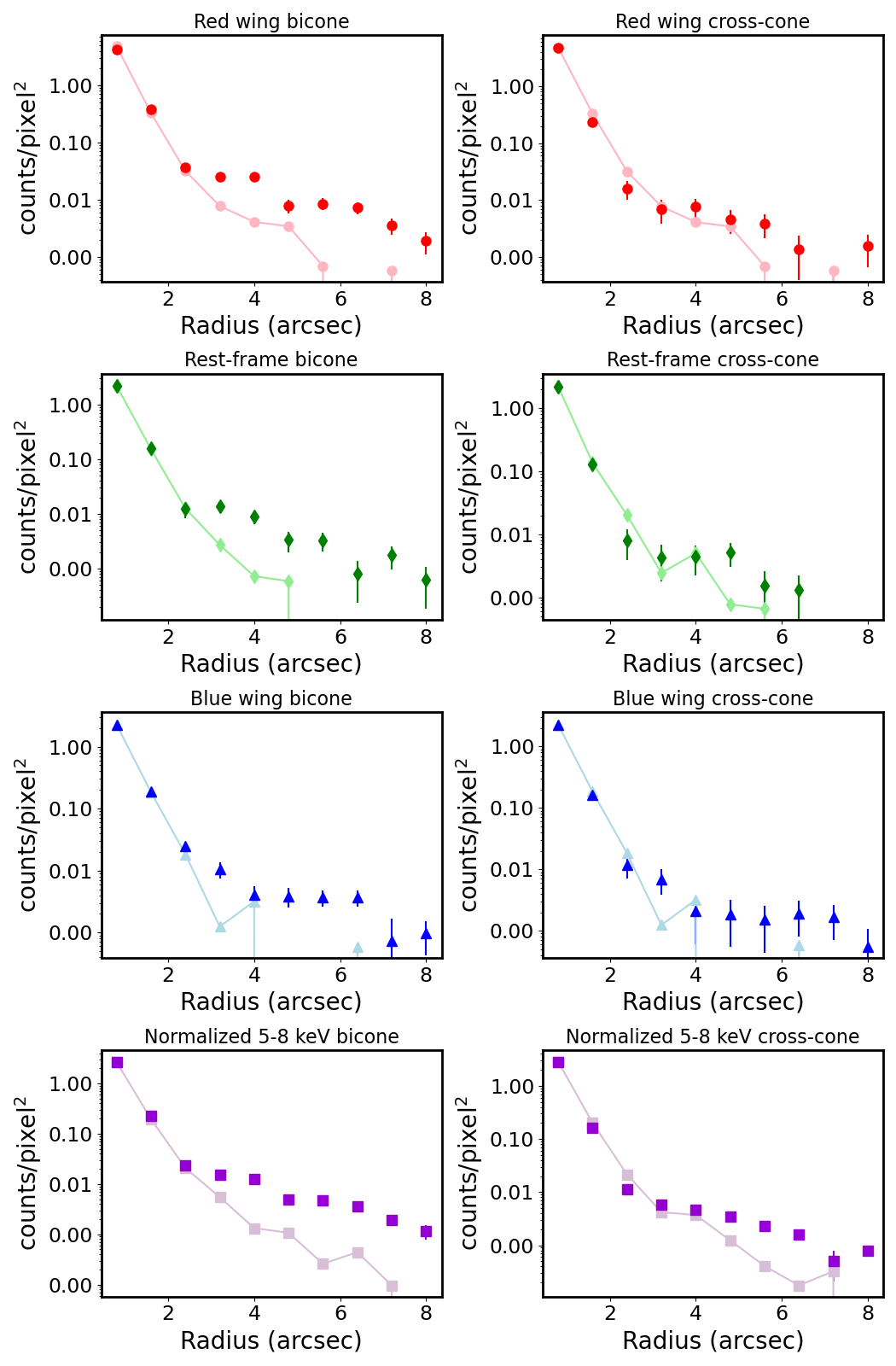}
\caption{Emission radial profiles for the red wing, rest-frame, blue wing, and normalized 5-8 keV continuum emission compared to that of the normalized \textit{Chandra} ACIS-S PSF. Bicone emission radial profiles are shown on the left, while cross-cone emission profiles are shown on the right. Counts and errors in the continuum band were normalized according to the difference between the model energy fluxes in each band. Each extraction bin was chosen to contain at least 10 counts in the red wing band. Note: pixel$^{2}$ indicates an element of area of an image (in this case, the pixel image is a square of 0.0615$''$ on the side = 1/8 of the ACIS pixel)}
 \label{fig:bicone_radial_profile} 
 \end{figure}

Fig. \ref{fig:bicone_radial_profile} shows the individual bicone emission profiles in the red and blue wings, rest frame, and continuum bands, compared to that of the \textit{Chandra} ACIS-S PSF normalized to the inner 0.5$''$, following the procedure used in \citetalias{trindadefalcao2023a}. The emission radial profiles were extracted with \textit{dmextract} from 1/8 subpixel images, with bin sizes varying to contain a minimum of 10 counts per bin in the red wing image dataset. In every instance, the observed emission exceeds that expected from the wings of the \textit{Chandra} ACIS-S PSF in the extended bicone region ($r>1.5''$). As detailed in Section \ref{sec:method}, our comprehensive discussion provides a strong basis for asserting that this excess emission is intrinsic to the bicone. This conclusion is further supported by the images in Figure \ref{fig:image}.

\begin{figure}[htb]
  \centering
 \includegraphics[width=18cm]{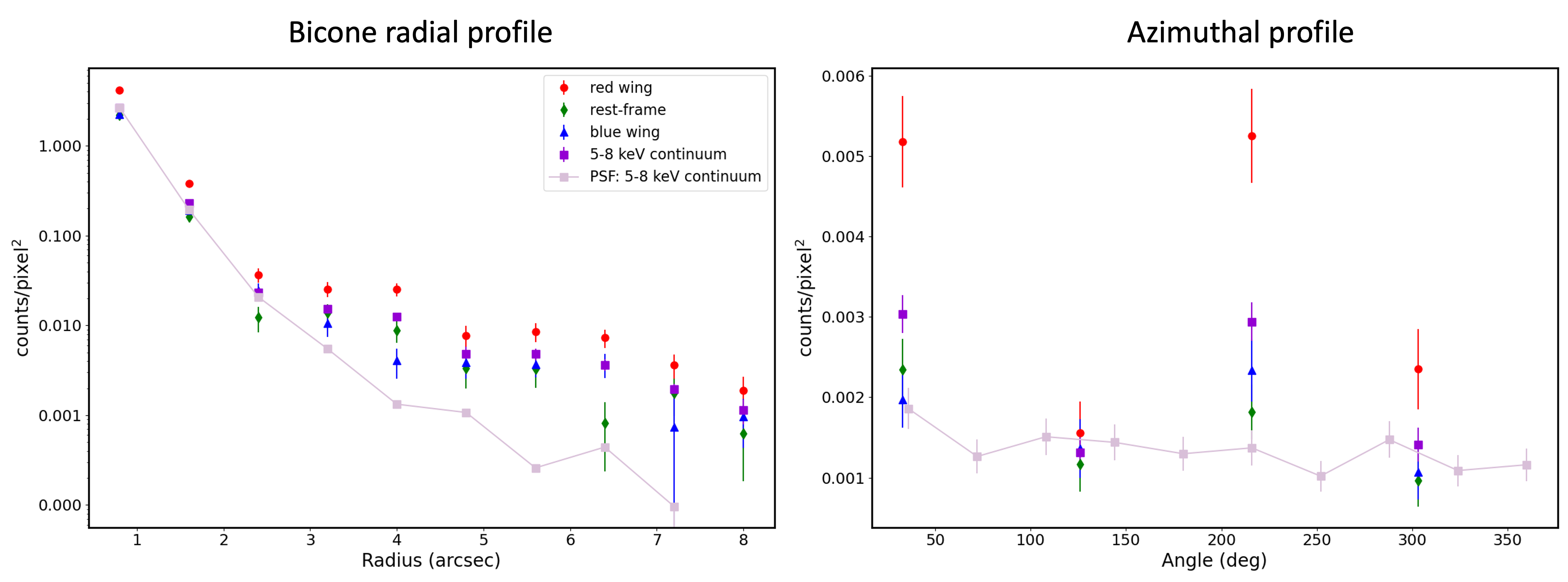}
\caption{\textbf{Left:} Bicone emission radial profiles for the red wing, rest-frame, blue wing, and normalized 5-8 keV continuum, compared to that of the \textit{Chandra} ACIS-S PSF normalized to the counts of the 5-8 keV continuum. Counts and errors in the continuum band were normalized according to the difference between the model energy fluxes in each band. For NGC 5728, 1$''$=200 pc, thus the limits of the plot range from 0 to 1 kpc. \textbf{Right:} Azimuthal emission profiles showing the angular dependence of the emission in the red wing, rest-frame, blue wing, and continuum bands. }
 \label{fig:radial_profile} 
 \end{figure}

 Fig. \ref{fig:radial_profile} (left panel) shows the bicone emission radial profiles on the left, and the azimuthal profiles on the right. The azimuthal profiles were extracted using a 1.5$''$–8$''$ annular region centered on the nucleus, with angular bins corresponding to the conical extraction regions shown in Fig. \ref{fig:ds9_regions}. All azimuthal profiles show significant bicone excesses, which is not expected from the azimuthally symmetrical \textit{Chandra} ACIS-S PSF (Section \ref{sec:method}). As in \citetalias{trindadefalcao2023a}, we determined the full extent of the emission by measuring the width at which the background-subtracted surface brightness from the emission radial profile (Fig. \ref{fig:radial_profile}, left) matches the background surface brightness in the same energy band \citep[e.g.,][]{fabbiano2017a, jones2020a, travascio2021a}. To ensure that our results were statistically significant, we binned the data into bins with a minimum significance of 3$\sigma$. The red wing is extended out to $\sim$6$''$ (1.2 kpc), while the blue wing and rest-frame emissions extend out to $\sim$5$''$ (1 kpc) radii. The line emission (narrow-line: 70 counts in the biconical region; red: 120 counts in the biconical region; blue: 73 counts in the biconical region) surpasses the emission of the normalized hard continuum (5-8 keV, 32 counts in the biconical region), for 1.5$''<r<8''$, by 6.5$\sigma$, 5.4$\sigma$ and 3.7$\sigma$, respectively. Non-nuclear point sources were, in all cases, excluded from the radial and azimuthal profile analysis, as shown in Fig. \ref{fig:ds9_regions}. Even considering a factor of 20\% uncertainties in the calibration of the wings of the \textit{Chandra} ACIS-S PSF at these radii (\citealt[][Fig. 11]{jerius2002a}; see Appendix of \citealt{ma2023a}) the extent is highly significant, as it is more than 1 order of magnitude higher than the PSF (see Fig. \ref{fig:radial_profile}, left). Instead, the azimuthal emission profiles (Fig. \ref{fig:radial_profile}, right) show that the cross-cone emission is consistent with the \textit{Chandra} ACIS-S PSF within statistical uncertainties, in all energy bands considered.

\medskip

In summary, the \texttt{marx} model PSF provides an adequate representation of the \textit{Chandra} ACIS-S PSF, even in the wings, for our purposes. We based this conclusion on:
(1) The comparison of bicone and cross-cone emission radial profiles, which provides a self-calibration in the assumption of lack of extent in the cross-cone direction; (2) The agreement between the \texttt{marx} model PSF and the cross-cone 4-7 keV surface brightness profiles \citepalias{trindadefalcao2023a}; (3) The comparison with the empirical PSF provided by PKS 1055+201 (see above and Appendix \ref{sec:appb});
However, note that localized $\sim$20\% uncertainties in the \texttt{marx} calibration certainly remain in the outer core, and a $\sim$5\% spurious feature is observed in the inner core (see \citealt{ma2023a}, Appendix A).

We also compared the \texttt{marx} model PSF with the ACIS-S PSF wings at large radii, using the CXC calibration of the wings with Her X-1 observations (Appendix \ref{sec:appc}). This comparison shows that the \texttt{marx} model PSF in the 4-7 keV energy band is valid within 5$''$ and underestimates the ACIS-S PSF wings by an increasing factor of up to $\sim$6 at 10$''$. Given the statistics of our data in these outer radii, these differences are within the margin of error.

\section{Discussion}
\label{sec:disc_conc}

In this Section, we first summarize the results of our spectral and spatial analysis, and their implications for the existence of the Fe \ka wings (Section \ref{sec:robs_disc}). We then discuss possible physical emission scenarios, such as fluorescence (Section \ref{sec:fluorescence}) and shocked emission from the interaction with the host ISM (Section \ref{sec:ufos}). Finally, in Section \ref{sec:implications}, we discuss the implications of these results to the overall picture of AGN feedback and outflows.

\subsection{Robustness and Spatial Properties of the Spectral Wings} 
\label{sec:robs_disc}

In this paper, we reanalyzed the \textit{Chandra} ACIS-S data of NGC 5728, confirming the existence and determining the properties of the newly discovered broad spectral wings to the 6.4 keV Fe \ka line. These features were first detected in the spectral analysis of the extended bicone emission (1.5$''$-8$''$, 300-1,600 pc in the bicone – see Figure \ref{fig:ds9_regions} and \citetalias{trindadefalcao2023a}).

We first fitted the full 0.3-8 keV band spectral data with a simple power-law plus Gaussians. This fit yielded significant excesses, to the red (5.4$\sigma$) and blue (3.7$\sigma$) of the neutral Fe \ka 6.4 keV line (Section \ref{sec:pheno}). To test the robustness of these wings to different and more complex spectral models, we employed multi-component spectral models, including photoionization (\texttt{CLOUDY}), thermal emission models (\texttt{APEC}), and the slab reflection model \textit{\texttt{xspexmon}}, to model the hard ($>$3 keV) continuum and 6.4 keV neutral line (Section \ref{sec:physicalmodels}). We also employed \texttt{CLOUDY} + \texttt{APEC} only models, to model the high energy spectrum; both models can reproduce the hard continuum, and \texttt{CLOUDY}, with a slab geometry, can also model reflection spectra. We ensured that all these complex models gave an adequate representation of the soft ($<$3 keV) line-dominated spectra by excluding models with consistent correlated residuals in the soft energy band. As summarized in Section \ref{sec:summary_spec}, we found that the red wing is robust, appearing as a $\sim$3$\sigma$ or larger excess in virtually all 33 fits, while the blue wing may be in part explained by high ionization Fe line emission but still leaving $\sim$2$\sigma$ or larger residuals. 

We also excluded that the red wing could arise from a Compton shoulder (CS) emission, both by comparing its properties to CS flux and shape (Section \ref{sec:CS}) and by spectral fits with a suite of models used in the study of the X-ray spectra of CT AGNs, which include CS calculations. All these fits failed to reproduce the observed red wing. The red wing emission is notably broader than would be expected from an unresolved emission-line with $E_{\rm max}$=6.0 keV and $E_{\rm min}$=5.5 keV. From the point of view of statistical significance, the blue wing is marginal but is not fully explained by any of the many models fitted to the NGC 5728 extended-spectrum.\par 

Note that the emission under analysis here is spatially extended (Section \ref{sec:images}) and not connected with the nuclear AGN emission (as already reported in \citetalias{trindadefalcao2023a}, for both hard continuum and host rest frame 6.4 keV emission). We separately investigated the spatial properties of the emission in both the red and blue wings and in the rest frame line and adjacent continuum by comparing them with the \textit{Chandra} ACIS-S PSF in their relevant energy bands. In all cases, the emission is more extended than the PSF wings in the direction of the bicone, while it is consistent with the PSF in the cross-cone direction. \textit{Azimuthal dependencies are not expected in the \textit{Chandra} PSF at the aim point.} We obtained an empirical 4-7 keV PSF from archival \textit{Chandra} observations of the quasar PKS 1055+201 and demonstrated that it is consistent with the \textit{Chandra} ACIS-S PSF model and the cross-cone radial dependence of NGC 5728 while greatly under-predicting the bicone emission. Appendix \ref{sec:appb} gives a detailed discussion of this empirical PSF and its comparison with \texttt{marx} ACIS-S PSF models. The hard continuum and Fe \ka complex exhibit clear spatial extent, extending from 1.5$''$ to approximately 5$''$-6$''$ ($\sim$1 kpc) across all energy bands within the full range (0.3-8 keV), excluding any potential origin from the nuclear torus. 

Contamination from the bright nuclear region is unlikely since the PSF fraction at $\sim$6 keV in these conical regions is only $\sim$5\%, while the red and blue wings have approximately 20\% of the continuum flux. A comparison of the bicone and nuclear spectra (see also \citetalias{trindadefalcao2023a}) excludes the spectral Fe \ka wings as being the result of PSF spillover: although both the red and blue wings may be present in the nuclear spectrum (as reported in \citetalias{trindadefalcao2023a}), the emission in these energy bands is much more prominent in the bicone. In the nucleus, the EW$_{\rm line}$/EW$_{\rm Fe~Ka}$ values for the red and blue wings are 0.3 $\pm$ 0.1 and 0.1 $\pm$ 0.1 respectively, increasing to 0.7 $\pm$ 0.2 and 0.8 $\pm$ 0.3 in the bicone, for the red and blue wings, respectively. The same trend is observed in the model energy flux, where the E$_{\rm line}$/E$_{\rm Fe~Ka}$ values are 0.2 $\pm$ 0.1 and 0.2 $\pm$ 0.1 in the nucleus, and 0.7 $\pm$ 0.3 and 0.8 $\pm$ 0.2 in the bicone. Since the nuclear and bicone values are different, this comparison` rules out contamination due to PSF spillover from the nuclear region. Moreover, any PSF spillover of a source at the aim point would not have the strong biconical azimuthal feature observed in NGC 5728. 

The spatial properties of the extended Fe \ka complex and the lack of nuclear contamination strongly associate this emission with processes in the ISM of the ionization bicone. This extended Fe \ka complex emission is aligned with the direction of the extended soft X-ray emission (\citetalias{trindadefalcao2023a}), optical line emission, and the radio jet \citep[][Figure 5 therein]{durre2018a}. Similar spatially extended Fe \ka wings have been recently reported in the Seyfert 2 AGN Mrk 34 \citep{maksym2023a}. The emission features observed in NGC 5728 and described here exhibit significantly higher significance and larger spatial extent.

\subsection{Fluorescent Relativistic Winds} 
\label{sec:fluorescence}

As discussed in Section \ref{sec:introduction}, \textit{Chandra} ACIS-S observations have provided widespread evidence of extended non-nuclear hard continuum and Fe \ka emission in CT AGNs (see review, \citealt{fabbiano2022b}). In the central regions of the Milky Way, individual fluorescing molecular clouds, remnants of past Sgr A* activity, have been studied with \textit{Chandra} and \textit{XMM-Newton} \citep[e.g.,][]{ponti2015a}. In the CT AGN ESO 428-G014 \citep{feruglio2020a,fabbiano2018a}, comparison with \textit{ALMA} data has shown a close correspondence between the hard X-ray emission, the Fe \ka line, and the spatial distribution of the molecular clouds, strengthening the reflection and fluorescence scenario. In the extended bicone of NGC 5728, the large EW of the red wing (EW=1.8 keV $\pm$ 0.4) and blue wing (EW=2.6 keV $\pm$ 0.3) suggest that these emissions may arise from reflection by the \ka transition of neutral Fe species, possibly found in molecular clouds, dust, or in a moderately ionized X-ray wind.

In Mrk 34, assuming that the observed red and blue wings of the rest-frame fluorescence 6.4 keV Fe \ka are due to Doppler-shifted fluorescent emission, \citet{maksym2023a} suggested the presence of strong winds, with line-of-sight velocities $v\sim$15,000 km s$^{-1}$, within $\sim$200 pc of the nucleus. In NGC 5728, if the wings emission is indeed due to redshifted and blueshifted neutral Fe \ka, then the implied velocities along the line-of-sight would be $v\sim$0.06$c$-0.14$c$ for the red wing and $v\sim$-0.09$c$ for the blue wing, consistent with a symmetric outflow. Assuming the \citet{durre2019a} bicone outflow model (with an inclination to the line-of-sight of $i$=47$\degree$, and the bicone axis nearly parallel to the plane of the galaxy) and considering biconical symmetry, the deprojected velocities for the red and blue wings would be $v_{\rm deproj}\sim$0.08$c$-0.2$c$ (redshifted) and $v_{\rm deproj}\sim$-0.13$c$ (blueshifted), respectively.

Given the inclination of the NGC 5728 bicone, if these neutral Fe \ka winds are associated with expanding biconical outflows \citep[e.g.,][]{fischer2013a}, then it is expected that the red and blue wings will be similarly extended along the bicone (see \citealt{maksym2023a}, Fig. 5), as observed within statistics in NGC 5728 (Section \ref{sec:images_subsection}, Figs. \ref{fig:image} and \ref{fig:radial_profile}). Therefore, in this scenario, the blue wing emission would be comparable to the red wing, and the blue wing therefore should not be dominated by the emission of highly ionized Fe lines. The blue wing yields $\sim$60\% of the counts detected in the red wing (see Fig. \ref{fig:image}). Within statistics, this is consistent with the factor of $\sim$2 reduction in the HRMA+ACIS-S effective area from 6 to 7 keV\footnote{https://cxc.harvard.edu/proposer/POG/html/chap6.html\#tth\_sEc6.5 (Fig. 6.5 therein)}.

\subsection{Shocked Emission from UFO Winds Interacting with the Host ISM}
\label{sec:ufos}

A different possibility is that the blue wing may arise from highly ionized and extended Fe XXV/Fe XXVI emission, with line-of-sight velocities of $v\sim$10,000 km $s^{-1}$ or $v\sim$-0.03$c$, extending out to $\sim$1 kpc from the AGN. These velocities correspond to deprojected velocities of $v_{\rm deproj}\sim$-0.04$c$, consistent with the velocity range associated with UFOs. UFOs are commonly identified through high-excitation Fe XXV and Fe XXVI absorption lines in the hard X-ray band (7–10 keV; see e.g., \citealt{tombesi2010a} and references therein). These lines are the strongest absorption lines produced at high ionization. High ionization is expected for winds originating in the innermost hot regions of the accretion disk \citep[e.g.,][]{laha2021a}. If the extended blue emission indeed originates from highly ionized, relativistic Fe XXV+Fe XXVI outflows, it may indicate that these lines form \textit{within the UFO itself}. The host rest-frame neutral Fe \ka emission might be produced at the shock front \citep[e.g.,][]{travascio2021a} or could be associated with independent fluorescent emission from molecular clouds in the host galaxy. 

In this scenario, the observed redshifted emission (red wing) may arise from gas streaming back along the edges of the shock, with similar (or lower) velocity to that of the shocking outflow, assuming momentum conservation, and producing a blend of redshifted Fe XXII-Fe XXIV lines. Instead, the red wing observed line-of-sight velocity is $v\sim$0.06$c$-0.25$c$, significantly higher than that of the blue wing. 

\begin{figure}[htb]
  \centering
 \includegraphics[width=18cm]{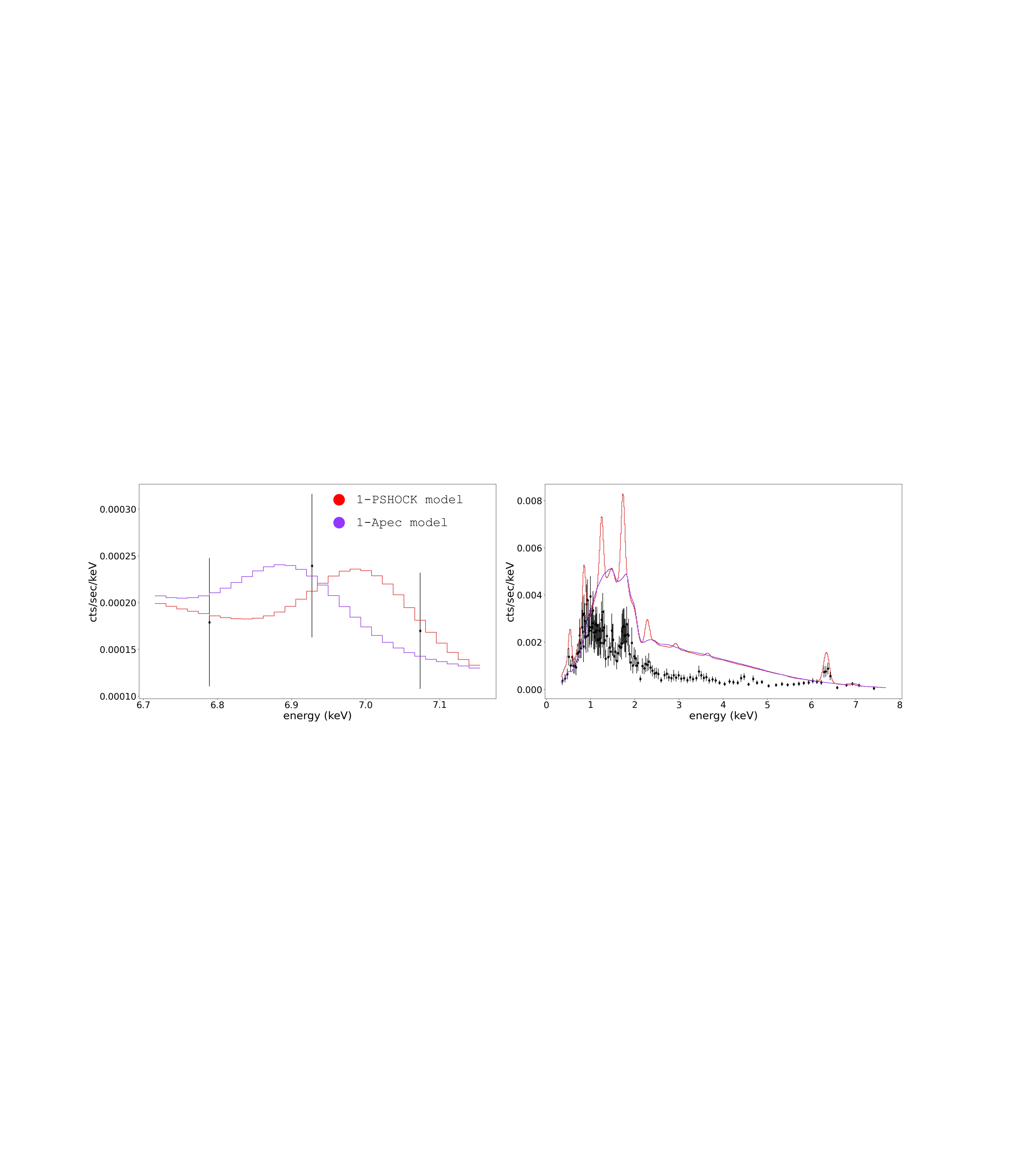}
\caption{\textit{Chandra} ACIS-S extended bicone spectrum in NGC 5728. We fit the emission in the blue wing (6.5-7.5 keV) as shocked emission, arising from highly ionized Fe lines. \textbf{Left:} We used single-component \texttt{xspshock} and \texttt{xsapec} to model the 6.5-7.5 keV spectrum, forcing these models to fully fit the emission in the blue wing. \textbf{Right:} The continuum predicted by these models greatly exceeds that of the observed low energy spectrum. To fully fit the blue being emission, \texttt{xspshock} overpredicts the neutral Fe line emission (\texttt{xsapec} does not model neutral fluorescence). Both models greatly overpredicts the emission in the soft band. } 
\label{fig:pshock}
 \end{figure}

Another problem with this scenario is that modeling the emission in the blue wing as shocked emission yields a continuum that is inconsistent with the observed spectrum. Fitting the extended bicone in the 6.6-7.5 keV band with a 1-component \texttt{APEC} or 1-component \texttt{PShock}\footnote{This model assumes a plane-parallel shocked plasma with constant post-shock ion and electron temperature, ionization timescale, and element abundances, providing a useful approximation for all cases in which X-ray emission is produced in a shock front.} \citep{borkowski_2001a} model, and forcing the model to fully fit the emission in the blue wing extremely overpredicts the continuum at energies $<$6 keV (Fig. \ref{fig:pshock}), due to the insufficient production of Fe XXV and Fe XXVI in a high temperature (kT$>>$ 20 keV) plasma.

\subsection{UFOs, BALs, and the High-Velocity NGC 5728 Winds}
\label{sec:implications}
Regardless of the emission scenario, the velocities inferred for the red and blue wings in NGC 5728 fall within the range of UFOs \citep[e.g.,][]{tombesi2010a}. UFOs have so far been detected in X-rays as absorption features with velocities of $v\sim$0.03$c$-0.3$c$, frequently associated with highly ionized Fe XXV or Fe XXVI species \citep[e.g.,][and also the P-Cygni– like profile in \citealt{nardini2015a}]{tombesi2010a, gofford2013a, tombesi2014a, chartas2021a}. 

In NGC 5728, neutral Fe \ka is a more likely explanation. The production of such a significant amount of Fe \ka fluorescence suggests a large column density (log N$_{H}\geq$23, $\tau_{\rm Compton}>0.1$), indicating a likely association with dusty molecular material that may be accelerated through ablation off molecular clouds \citep{maksym2020a, maksym2021a, travascio2021a}. 

A key challenge to this scenario is the kpc-scale location of the high-velocity winds found in NGC 5728. UFOs are much more compact [typically $<$0.03 pc, \citet{tombesi2012a}, or, in some models, even on accretion disk scales, \citet{gallo2011a, gallo2023a}].

A closer analog to the discovered extended, high-velocity emission may be high-ionization BALs \citep{arav2018a, arav2020a, miller2020a}. BALs are often observed in AGN spectra as absorption features of high ionization species, such as C~IV, O~VI, N~V, and S~IV. BALs are found in approximately 10\%-30\% of quasars \citep{netzer2013a}. Utilizing FUV density diagnostics, \citet{arav2018a} measured BALs distances exceeding 100 pc in half of the BAL quasars in their sample, and found 12\% of their sample located at distances exceeding 1 kpc. The column density of the slab of gas must be roughly CT to produce the detected neutral Fe emission, which is plausible for BALs given their X-ray weakness \citep{grupe2003a,gibson2009a}.

\citet{serafinelli2019a} have shown the coexistence of three distinct absorber types in the quasar PG 1114+445, suggesting a multiphase and multiscale outflow. In this case, the outflow starts off as a UFO, which subsequently entrains gas from the ISM, resulting in an "extended UFO" at $\sim$100s pc. The extended UFO maintains high velocities but exhibits lower ionization and column density. This wind subsequently decelerates to a WA at a distance of $\sim$1 kpc from the AGN. Our observations of NGC 5728 may be analogous to this scenario, albeit without the deceleration at larger distances. 

\bigskip

Models for effective AGN-ISM feedback require a minimum of 5\% of the AGN bolometric luminosity to effectively unbind the host ISM \citep{dimatteo2005a}, while disrupting molecular clouds to stop star formation could be achieved with only 0.5\% \citep{hopkins2010a}. Outflowing AGN winds, usually observed through optical spectroscopy, show large gas masses and velocities $v\sim$100-1000 km~s$^{-1}$, but do not seem to carry sufficient kinetic power to disrupt the host ISM (see review by \citealt{crenshaw2012a}). 

However, the kinetic power carried by AGN winds into the host galaxy rises rapidly with the radial velocity of the outflows, i.e., $L_{\rm KE}\propto v_{r}^{3}$. Therefore, the newly discovered semi-relativistic wings in NGC 5728 could carry $100^{3}$ ($\sim$ one million) times the kinetic power of optical [O~III] outflows for equal mass in the two phases, potentially becoming the dominant driver of AGN feedback in the local Universe. 
\section{Conclusions}
\label{sec:conclusions}

We have conducted a spectral and spatial analysis of the red and blue wing features present in the X-ray spectrum of the extended non-nuclear bicone emission ($\sim$5$''$, $\sim$1 kpc) of NGC 5728, first noted in \citetalias{trindadefalcao2023a}. This analysis shows that the spectral wings are robust over a suite of 33 spectral models and are extended beyond the nuclear AGN point source at high statistical significance. Since the red and blue wing fluxes are comparable in flux then, if attributed to redshifted and blueshifted neutral Fe \ka, the implication is that fluorescing material is moving away from the nucleus symmetrically with velocities $\sim$0.1$c$. These velocities are $\sim$100 times higher than those detected in optical emission lines \citep{durre2019a}, which are typical of biconical outflows in nearby AGNs \citep[e.g.,][]{fischer2018a}.

This work, together with \citetalias{trindadefalcao2023a}, shows that the bicones contain a multi-phase medium including hot ionized gas and a colder scattering medium, with a high velocity component. This newly discovered extended high velocity emission suggests a connection to the sub-parsec scale UFOs and to kpc-scale BALs, usually observed in absorption, but here seen in emission. 

We have considered a second scenario connected with highly ionized Fe \ka lines resulting from shocks in the ISM caused by an outflowing UFO. However, this scenario would produce a puzzling difference in the red and blue wings outflow velocities. Moreover, spectral modeling of the blue wing as entirely due to Fe~XXV, Fe~XXVI returns a continuum flux strongly in excess of the observations.

As the kinetic power associated with mass outflows increases as the cube of the velocity, this high velocity material may dominate the kinetic power of the wind and be capable of perturbing the host ISM, facilitating effective AGN feedback.

Given the limited statistics of our data the question of the detailed location and structure of these high velocity outflows within NGC 5728 remains unresolved. To provide more stringent constraints leading to a more comprehensive understanding of the origins of these extended fast winds, new and deeper \textit{Chandra} ACIS-S observations are required. \textit{HST} and \textit{JWST} data could help map the extended high ionization regions and determine their kinematics in detail.



\begin{acknowledgments}
This work was partially supported by NASA contract NAS8-03060 (CXC) and the \textit{Chandra} Guest Observer program grant GO0-21094X (PI: Fabbiano). The NASA ADS bibliography service was used in this work. We used the NASA/IPAC Extragalactic Database (NED), which is operated by the Jet Propulsion Laboratory, California Institute of Technology, under contract with the National Aeronautics and Space Administration. For the data analysis, we used the \texttt{CIAO} toolbox, \texttt{Sherpa}, and \texttt{DS9}, developed by the Chandra X-ray Center (CXC). This work used the photoionization code \texttt{CLOUDY}, the thermal code \texttt{xsapec}, and the shocked model \texttt{xspshock}. The spectral fitting models \texttt{xspexmon}, \texttt{MYTorus}, and \texttt{borus02} were also employed in our analysis. This work was initiated/performed in part at the Aspen Center for Physics, which is supported by National Science Foundation grant PHY-2210452. This work made use of the \textit{Chandra} Source Catalog. We gratefully acknowledge extensive and valuable conversations with the \textit{Chandra} X-ray Center (CXC) Calibration team, Vinay Kashyap, Diab Jerius, and Terry Gaetz, as well as with Tom Aldcroft of the CXC.
\end{acknowledgments}

\newpage 
\appendix

\section{Spectral fitting tables}
\label{sec:appa}

In this Appendix, we present the detailed spectral fit results that are summarized and discussed in Section \ref{sec:wings_definition}.

\setcounter{table}{0}
\renewcommand{\thetable}{A\arabic{table}}

\begin{table}[h]
\caption{\textbf{Phenomenological models} spectral fitting results for the extended 0.3-8 keV bicone spectrum.}
\label{tab:pheno_models}
\begin{center}       
\begin{tabular}{|c|c|c|c|} 
\hline
  \textbf{Model} & \textbf{$\chi^{2}_{\nu}$} (d.o.f) & \textbf{\textit{\texttt{softpowerlaw}}} & \textbf{Significance [red, blue] wings}\\
  \textbf{description}& & ($\Gamma$)& \\
    \hline
    \hline  

 without [red, blue] wings& 0.68 (132)& 1.3$\pm$0.6&  [5.4$\sigma$, 3.7$\sigma$]\\
 \hline
 with [red, blue] wings& 0.42 (138)& 1.4$\pm$0.4&  [0.7$\sigma$, 0.5$\sigma$]\\

  \hline
  \hline
 
  \underline{\textbf{Emission Lines:}} & & &  \\
  Line Energy & Line Flux & Line Width & Identification \\
 (keV)& (10$^{-6}$ ph~cm$^{-2}$~s $^{-}$)& (keV) & \\
 \hline

  0.5$\pm$0.1 & 3.5$\pm$2.9 & 0.13&$^{a}$O~VII \\
  \hline
  
  0.8$\pm$0.1 &10.2 $\pm$3.4 &0.22&$^{a}$O~VIII, $^{a}$Fe~XVII blend \\ 
  \hline

  1.3 $\pm$0.2 &0.3 $\pm$0.1 &0.11&Mg~XI  \\ 
  \hline 

  1.8$\pm$0.2 &0.5 $\pm$0.1 &0.11 &Si XIII \\ 
  \hline

  2.3$\pm$0.1 &0.3 $\pm$0.1 &0.12&S \ka, S XV blend  \\ 
  \hline

  3.5&<0.1 & 0.15&Ar XVII, Ca \ka blend \\ 
  \hline

  4.5&<0.1&0.15&Ca \ka  \\ 
  \hline

  6.0$\pm$0.2 &0.8 $\pm$0.3 & 0.26&$^{**}$red wing  \\ 
  \hline

  6.4$\pm$0.1&1.0 $\pm$0.2 & 0.10&neutral Fe \ka  \\ 
  \hline

  7.0$\pm$0.3 &0.7 $\pm$0.2 & 0.12&$^{**}$blue wing \\ 
  \hline
\end{tabular}
\end{center}
\centering
\footnotesize$^{a}$ Lines are blended in the ACIS-S spectrum $<$ 1.3 keV. These are tentative identifications.\\
\footnotesize$^{**}$ These components were not included in the first set of phenomenological models described in Section
\ref{sec:pheno}.
\end{table}

\begin{table}
\begin{center}
\caption{\textbf{\textit{\texttt{xspexmon+softpowerlaw+CLOUDY+APEC}}} spectral fitting results for the extended 0.3-8 keV bicone spectrum.}
\label{tab:models}

\begin{tabular}{|c|c|c|c|c|c|} 
    \hline
    \textbf{Model} & \textbf{$\chi^{2}_{\nu}$} (d.o.f.) & \textbf{log N$_{H_{\rm slab}}$} & log $U$ & k$T$ & \textbf{Significance} \\
  \textbf{description}  & & (cm$^{-2}$)&  &(keV) & \textbf{ [red, blue] wings}\\ 
    \hline
    \hline    
    \textit{\texttt{xspexmon}}+2\texttt{gauss}+\textit{\texttt{softpowerlaw}}+& 0.85 (136)&  20.7$\pm$0.6 &1.6 $\pm$0.8&0.7$\pm$0.2  & [3.3$\sigma$, 2.1$\sigma$] \\
   (1+1)  \texttt{CLOUDY}+\texttt{APEC} models&   &  &    & & \\
    \hline

    \textit{\texttt{xspexmon}}+2\texttt{gauss}+\textit{\texttt{softpowerlaw}}+& 0.87 (136)& 20.7$\pm$0.6 &1.5 $\pm$0.3&0.8$\pm$0.2  &  [3.8$\sigma$, 2.3$\sigma$] \\
   (2+1)  \texttt{CLOUDY}+\texttt{APEC} models&  & 23.0$\pm$0.6 & -1.3$\pm$1.2 &  &   \\
    \hline

    \textit{\texttt{xspexmon}}+2\texttt{gauss}+\textit{\texttt{softpowerlaw}}+& 1.04 (133)& 20.1$\pm$0.6 &1.5$\pm$0.2 &0.9$\pm$0.4& [3.8$\sigma$, 2.2$\sigma$] \\
   (3+1)  \texttt{CLOUDY}+\texttt{APEC} models&  & 23.3$\pm$0.2& -1.8$\pm$0.8 &  &   \\
    &   & 21.2$\pm$0.4 & -0.6$\pm$0.2 &  &  \\
   \hline
   
    \textit{\texttt{xspexmon}}+2\texttt{gauss}+\textit{\texttt{softpowerlaw}}+& 0.78 (134)& 20.3$\pm$1.1&1.6$\pm$0.4 &0.7$\pm$0.4  & [4.0$\sigma$, 2.3$\sigma$] \\
   (2+2)  \texttt{CLOUDY}+\texttt{APEC} models&  &23.2$\pm$0.4 & -1.6$\pm$1.1 & unconstrained &   \\
    \hline

    \textit{\texttt{xspexmon}}+2\texttt{gauss}+\textit{\texttt{softpowerlaw}}+& 0.75 (137) & 20.1$\pm$1.1 &1.6$\pm$0.3 &1.4$\pm$1.1  & [3.3$\sigma$, 2.1$\sigma$] \\
   (1+2)  \texttt{CLOUDY}+\texttt{APEC} models&   &  & & 0.7$\pm$0.2 &    \\
    \hline

     \textit{\texttt{xspexmon}}+2\texttt{gauss}+\textit{\texttt{softpowerlaw}}+& 0.73 (135) & 20.2$\pm$1.2 &1.4$\pm$0.9 &0.7$\pm$0.2  & [3.2$\sigma$, 2.2$\sigma$] \\
   (1+3)  \texttt{CLOUDY}+\texttt{APEC} models& &   &  & 1.5$\pm$0.4 &    \\
    &   & &  & unconstrained &   \\
   \hline
   \hline

\end{tabular}
\end{center}
\centering
\footnotesize$^{**}$$\Gamma_{\rm soft}$ and $\Gamma_{\rm ref}$ are unconstrained in all cases.
\end{table} 

\begin{table}
\begin{center}
\caption{\textbf{\textit{\texttt{xspexmon+CLOUDY+APEC}}} spectral fitting results for the extended 0.3-8 keV bicone spectrum.}
\label{tab:modelsII}

\begin{tabular}{|c|c|c|c|c|c|} 
    \hline
    \textbf{Model} & \textbf{$\chi^{2}_{\nu}$} (d.o.f.) & \textbf{log N$_{H_{\rm slab}}$} & log $U$ & k$T$ & \textbf{Significance} \\
  \textbf{description}  & & (cm$^{-2}$)&  &(keV) & \textbf{ [red, blue] wings}\\ 
    \hline
    \hline  
 \textit{\texttt{xspexmon}}+2\texttt{gauss}+& 0.97 (139)  &21.5$\pm$0.3 &1.4$\pm$0.3 &0.4$\pm$0.2  & [3.4$\sigma$, 1.8$\sigma$] \\
    (1+1)  \texttt{CLOUDY}+\texttt{APEC} models &  & & & &  \\
    \hline
    
   \textit{\texttt{xspexmon}}+2\texttt{gauss}+& 0.65 (136)&  20.7$\pm$0.9 & 1.5$\pm$0.4 & 0.9$\pm$0.3 & [3.9$\sigma$, 2.2$\sigma$] \\
  (2+1)  \texttt{CLOUDY}+\texttt{APEC} models &  & 23.1$\pm$0.9 & -1.1$\pm$1.1 &  & \\
    \hline

    \textit{\texttt{xspexmon}}+2\texttt{gauss}+& 0.66 (133)&  20.7$\pm$0.7 & 1.5$\pm$0.4 & 0.9$\pm$0.3 & [3.9$\sigma$, 2.2$\sigma$] \\
   (3+1)  \texttt{CLOUDY}+\texttt{APEC} models& &  23.2$\pm$0.8 & -1.3$\pm$0.7 &  & \\
   &  & 21.4$\pm$0.5 & -0.6$\pm$1.1 &  &  \\
    \hline

    \textit{\texttt{xspexmon}}+2\texttt{gauss}+& 0.75 (137)&  20.1$\pm$1.1 & 1.5$\pm$0.2 & 0.1$\pm$0.1 & [3.3$\sigma$, 1.9$\sigma$] \\
   (1+2)  \texttt{CLOUDY}+\texttt{APEC} models & &   &  & 0.7$\pm$0.4&  \\
   \hline
   
   \textit{\texttt{xspexmon}}+2\texttt{gauss}+& 0.74 (135)&  unconstrained & unconstrained & 0.7$\pm$0.3 & [3.3$\sigma$, 1.4$\sigma$] \\
   (1+3)  \texttt{CLOUDY}+\texttt{APEC} models&  & &  & 1.3$\pm$0.5&   \\
   & & &  & unconstrained &   \\
    \hline

     \textit{\texttt{xspexmon}}+2\texttt{gauss}+& 0.64 (134)&  20.6$\pm$0.7 & unconstrained & 0.2$\pm$0.1 & [3.3$\sigma$, 2.0$\sigma$] \\
   (2+2) \texttt{CLOUDY}+\texttt{APEC} models &  & 23.1$\pm$1.1 & unconstrained & 0.8$\pm$0.4&   \\
\hline
\end{tabular}
\end{center}
\centering
\footnotesize$^{**}$$\Gamma_{\rm ref}$ is unconstrained in all cases.
\end{table}

\begin{table}
\caption{\textbf{\texttt{CLOUDY/APEC}} spectral fitting results for the extended 0.3-8 keV bicone spectrum.}
\label{tab:red_wing_alternative} 
\begin{center}       
\begin{tabular}{|c|c|c|c|c|c|}
   
\hline
    \textbf{Alternative model}& \textbf{$\chi^{2}_{\nu}$} (d.o.f.) & 
  \textbf{log N$_{H_{\rm slab}}$} &\textbf{log U} & kT& \textbf{Significance} \\
    
    \textbf{description}& & (cm$^{-2}$) &  & (keV) & \textbf{ [red, blue] wings}\\ 
    \hline

    (1+1) \textbf{\texttt{CLOUDY}}+\textbf{\texttt{APEC}} models& 1.3 (141)& 20.3$\pm$1.2  & 1.7$\pm$0.2&0.3$\pm$0.1 & [4.6$\sigma$, 3.2$\sigma$] \\
    \hline

    (2+1) \textbf{\texttt{CLOUDY}}+\textbf{\texttt{APEC}} models& 0.7 (138)& 20.3$\pm$0.7  & 1.6$\pm$0.2&0.8$\pm$0.1 & [4.4$\sigma$, 2.6$\sigma$] \\
    & & 23.4$\pm$0.3&-1.1$\pm$0.7& &\\
    \hline

    (3+1) \textbf{\texttt{CLOUDY}}+\textbf{\texttt{APEC}} models& 0.7 (135)& 20.2$\pm$0.9  & 1.6$\pm$0.2&0.8$\pm$0.1 & [4.4$\sigma$, 2.6$\sigma$] \\
    & & unconstrained&-1.0$\pm$0.2& & \\
    & & 20.8$\pm$0.8&-0.3$\pm$1.2& & \\
    \hline

    (2+2) \textbf{\texttt{CLOUDY}}+\textbf{\texttt{APEC}} models& 0.7 (136)& 20.6$\pm$0.6  & 1.6$\pm$0.2&0.8$\pm$0.1 & [4.4$\sigma$, 2.2$\sigma$] \\
    & & 23.4$\pm$0.4&-1.1$\pm$0.6& 0.1$\pm$0.3& \\
   \hline
    
    (1+2) \textbf{\texttt{CLOUDY}}+\textbf{\texttt{APEC}} models& 1.1 (139)& 20.1$\pm$1.3  & 2.0$\pm$0.5&0.7$\pm$0.3 & [4.5$\sigma$, 3.7$\sigma$] \\
    & & && 0.1$\pm$0.3& \\
    \hline

    (1+3) \textbf{\texttt{CLOUDY}}+\textbf{\texttt{APEC}} models& 1.1 (137)& 20.1$\pm$1.3  & 2.0$\pm$0.5&0.7$\pm$0.3 & [4.6$\sigma$, 3.7$\sigma$] \\
    & & && 0.1$\pm$0.3& \\
    & & && unconstrained& \\
    \hline

    (2+2) \textbf{\texttt{CLOUDY}}+\textbf{\texttt{APEC}}+& 0.7 (134)& 20.3$\pm$0.9  & 1.6$\pm$0.2&1.4$\pm$0.1 & [4.5$\sigma$, 2.4$\sigma$] \\
    2\texttt{gauss} models& & 23.4$\pm$0.9&-1.1$\pm$1.2& 0.1$\pm$0.3& \\
   \hline
\end{tabular}
\end{center}
\centering
\end{table}

\begin{table}
\caption{\textbf{\textit{\texttt{xspexmon}}} spectral fitting results for the extended 3-8 keV bicone spectrum.}
\label{tab:red_wing_models}
\begin{center}       
\begin{tabular}{|c|c|c|c|}
    \hline
    \textbf{$\chi^{2}_{\nu}$} (d.o.f.) & \textbf{photon index} &  $\theta_{\rm inc}$ & \textbf{Significance} \\
     &($\Gamma$) & (degrees)  & \textbf{ [red, blue] wings} \\ 
    \hline
     
     1.14 (42)& 1.1$\pm$0.4 & 0 (frozen) & [3.1$\sigma$, 2.9$\sigma$] \\
    \hline

     1.17 (42)& 1.1$\pm$0.4  & 45 (frozen) & [3.2$\sigma$, 2.8$\sigma$] \\

    \hline

     1.41 (42)& 1.1$\pm$0.5 & 85 (frozen) & [2.9$\sigma$, 3.1$\sigma$] \\
    
    \hline

     1.17 (41)& 1.1$\pm$0.4  & unconstrained & [3.3$\sigma$, 2.8$\sigma$] \\
    \hline

      0.98 (41)& 1.1$\pm$0.6  & 45 (frozen) & [3.3$\sigma$, 3.0$\sigma$] \\
      & & Fe abund=unconstrained&\\
    \hline
\end{tabular}
\end{center}
\end{table} 

\begin{table}[h]
 \caption{\textbf{\texttt{MYTorus}} spectral fitting results for the extended 3-8 keV bicone spectrum.}
\label{tab:red_wing_mytorus}
\begin{center}       
\begin{tabular}{|c|c|c|c|c|}
\hline
    \textbf{$\chi^{2}_{\nu}$} (d.o.f.) & \textbf{photon index} & \textbf{N$_{\rm H_{\rm eq}}^{a}$} & \textbf{$\theta_{\rm inc}$} & \textbf{Significance} \\
     & ($\Gamma$) &(10$^{24}$cm$^{-2}$) & (degrees) & \textbf{ [red, blue] wings}\\ 
    \hline

      0.93 (40)& 1.5$\pm$0.6  & 4.0$\pm$10.1 &0 (frozen)  & [3.4$\sigma$, 2.1$\sigma$] \\
    \hline

    0.93 (40)& 1.4$\pm$1.3  & 2.99$\pm$34.9 &45 (frozen) & [3.4$\sigma$, 2.1$\sigma$] \\
    \hline

     0.90 (14)& 2.6$\pm$13.2 & 2.2$\pm$3.2 &90 (frozen) & [3.3$\sigma$, 1.9$\sigma$] \\
    \hline

     0.96 (39)& 1.4$\pm$5.3& 6.9$\pm$81.2&0.4$\pm$34.5 & [3.3$\sigma$, 2.1$\sigma$] \\
    \hline

\end{tabular}
\end{center}
\centering\footnotesize$^{a}$ N$_{H_{\rm eq}}$ is the equatorial column density
\end{table}

\begin{table}
\caption{\textbf{\texttt{borus02}} spectral fitting results for the extended 3-8 keV bicone spectrum.}
\label{tab:red_wing_borus02}
\begin{center}       
\begin{tabular}{|c|c|c|c|c|c|}
    
    \hline
    \textbf{$\chi^{2}_{\nu}$} (d.o.f.)& \textbf{photon index} & \textbf{log N$_{\rm H_{\rm tor}}^{a}$} & \textbf{cos($\theta_{\rm inc}$)} & \textbf{cos($\theta_{\rm tor}$)$^{b}$} & \textbf{Significance} \\

    & ($\Gamma$) & (cm$^{-2}$) & & &  \textbf{ [red, blue] wings}\\ 
    \hline

     0.77 (44)& 1.5$\pm$1.7 &23.9$\pm$1.6 & 0.95 (frozen) & 0.9$\pm$0.1&[3.2$\sigma$, 2.6$\sigma$] \\
    \hline

      0.76 (39)& unconstrained&unconstrained & 0.5 (frozen) & 0.8$\pm$0.3&[2.9$\sigma$, 2.5$\sigma$] \\
    \hline

      0.71 (44)& 1.6$\pm$2.5&23.9$\pm$0.2 & 0.05 (frozen) & 0.9$\pm$1.5&[2.9$\sigma$, 2.5$\sigma$] \\
    \hline

     0.73 (44)& unconstrained& 24.0$\pm$0.6 & 0.74$\pm$0.61 &0.9$\pm$0.5 &[2.9$\sigma$, 2.4$\sigma$] \\
    \hline
    0.72 (38)& unconstrained& 24.0$\pm$0.6 & 0.74$\pm$0.61 &0.9$\pm$0.5 &[2.9$\sigma$, 2.4$\sigma$] \\
     & & & Fe abund=0.8$\pm$0.6 & & \\
    \hline

\end{tabular}
\end{center}
\centering
\footnotesize$^{a}$ N$_{H_{\rm tor}}$ is the average column density of the torus;\\
\centering
\footnotesize$^{b}$ cos($\theta_{\rm tor}$) is the covering factor, where $\theta_{\rm tor}$ is the half-opening angle of the polar cutouts, measured from the symmetry axis toward the equator. \\
\footnotesize See https://sites.astro.caltech.edu/mislavb/download/ for more details.
\end{table} 

\clearpage
\section{Comparison between the \texttt{marx} PSF model and the empirical PSF from ObsID 7795 (PKS 1055+201)}
\label{sec:appb}
\setcounter{figure}{0}
\renewcommand{\thefigure}{B\arabic{figure}}

\begin{figure}[htb]
  \centering
 \includegraphics[width=18cm]{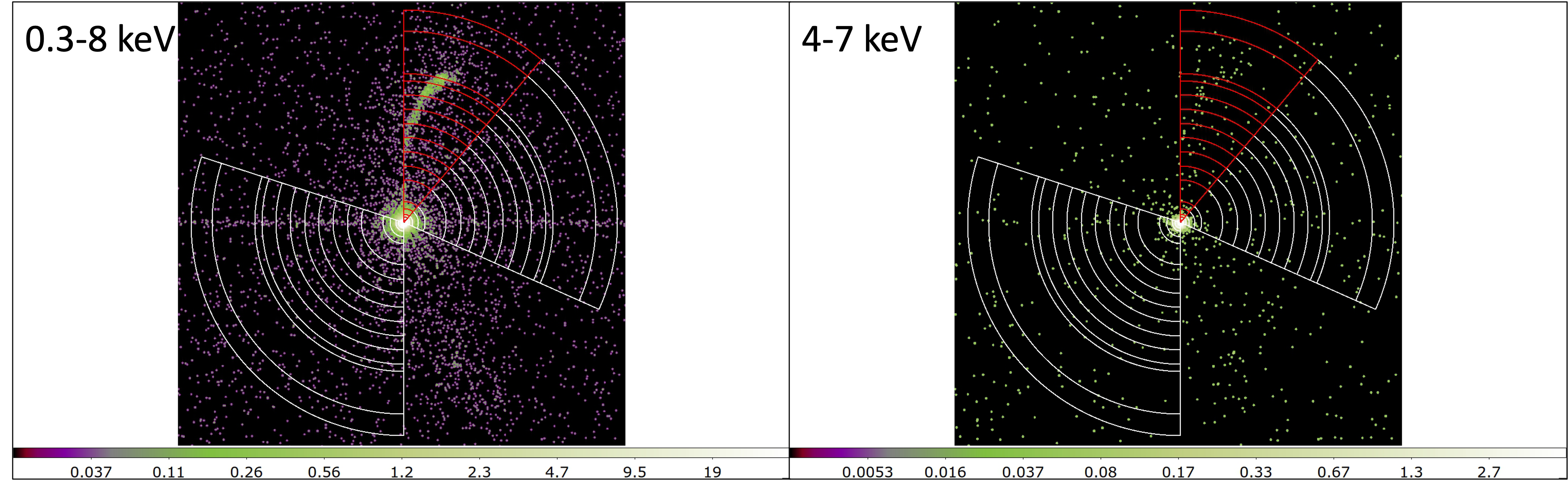}
\caption{Radial profiles of PKS 1055+201 for the 0.3-8 keV (left) and 4-7 keV (right) hard band. The background has been subtracted from the radial profiles using a 10$''$circular region.} 
 \label{fig:appb1} 
\end{figure}

Fig. \ref{fig:appb1} shows a wide-field image of PKS 1055+201 (0.3-8 keV, left; 4-7 keV, right). These images show both the central point-like source (see Fig. \ref{fig:pks}) and the extended X-ray counterpart of the radio jet \citep{schwartz2006a}, which is visible at radii $>$ 10$''$. To use PKS 1055+201 as an empirical PSF at large radii, we have derived bicone radial profiles by excluding the angular sector region shown in red in Fig. \ref{fig:appb1}, which comprises the X-ray jet. Extended X-ray emission associated with the radio jet may also contribute to the emission in the SE cross-cone region, which is also excluded. We have subtracted the field background by using a 10$''$ large off-source circular region.

\begin{figure}[htb]
  \centering
 \includegraphics[width=18cm]{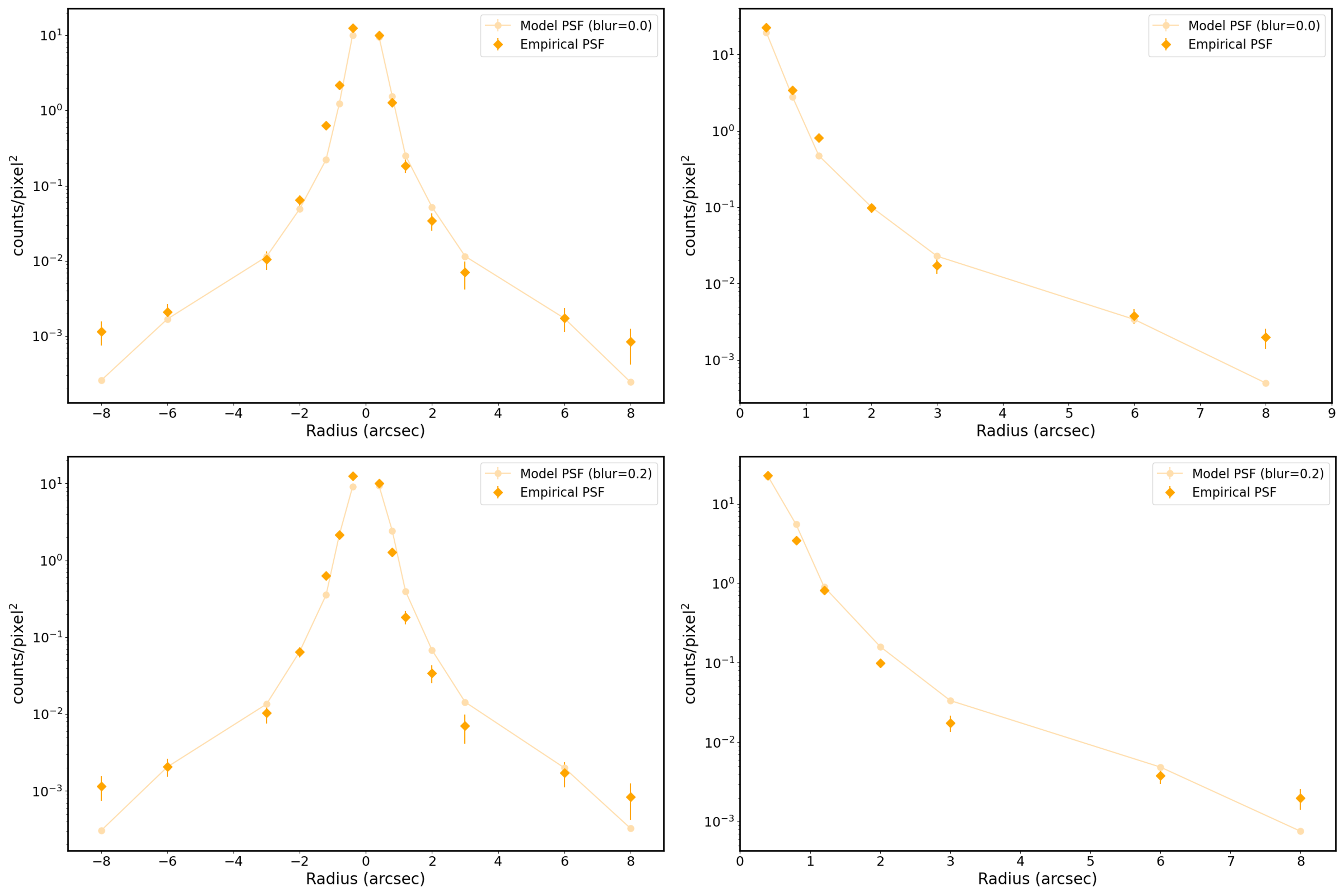}
\caption{Radial profiles of PKS 1055+201 for the 4-7 keV hard band (in orange) compared against the \texttt{marx} PSF (in light yellow), with different \textit{AspectBlur} (\textit{AspectBlur}=0$''$, top; \textit{AspectBlur}=0.2$''$, bottom), out to 8$''$. The background has been subtracted from the radial profiles using a 10$''$ circular region. The left column shows the radial profiles for the (NW-jet) cone (positive distances) and for the SE cone (negative distances), while the right column shows the total ((NW-jet)+SE) bicone emission.}
 \label{fig:appb2} 
 \end{figure}

Figure \ref{fig:appb2} compares the resulting PKS 1055+201 radial profile with the \texttt{marx} 4-7 keV PSF, with two blur factors: 0$''$ (top) and 0.2$''$ (bottom), within the central 8$''$ region. The latter was used in \citetalias{trindadefalcao2023a} and gives good agreement with the NGC 5728 in the central ($<$ 10$''$) and cross-cone profiles. The PKS 1055+201 radial profile is in remarkable agreement with the \texttt{marx} PSF model with aspect blur=0$''$, validating the \texttt{marx} PSF model and showing that a blur is not automatically introduced by the ACIS-S instrument.

We have also investigated possible pileup contributions to the central count distribution. Pileup would arise from two events with energies in the 2.0-3.5 keV range that would be detected as a single event in the 4-7 keV range. Within $r$=1.6$''$, we find 5,571 counts in the 2.0-3.5 keV range. The CXC pileup calculator estimates a 2\% pileup, i.e., 111 counts in this energy range, corresponding to 55 extra-counts between 4 and 7 keV. This is probably an overestimate because the 2.0-3.5 keV detection includes piled-up counts from lower energies. These counts (2\% of the detected 2,574 counts in the 4-7 keV range) would primarily occur in the centermost region. However, the ObsID 7795-\texttt{marx} model comparison shows that even considering only radii $>1''$ the count distribution of ObsID 7795 is narrower than a model built with \textit{AspectBlur}=0.2$''$.

Since the NGC 5728 data were co-added from 11 separate short observations \citepalias{trindadefalcao2023a}, we cannot exclude some blurring of the data resulting from statistical noise in the evaluation of the nuclear centroid. However, it is also possible that some real extended emission is also included in the central region, although this cannot be verified with the present data.

\begin{figure}[htb]
  \centering
 \includegraphics[width=18cm]{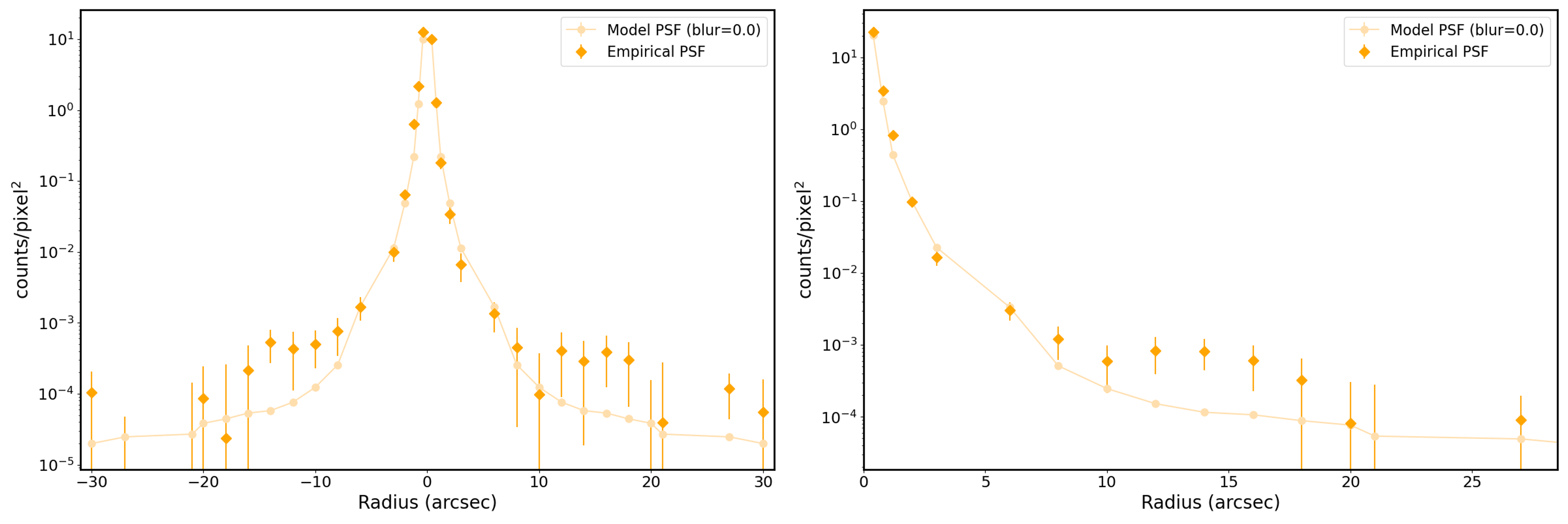}
\caption{Radial profiles of PKS 1055+201 for the 4-7 keV hard band (in orange) compared against the \texttt{marx} PSF (in light yellow), with \textit{AspectBlur}=0$''$, out to 30$''$. The background has been subtracted from the radial profiles using a 10$''$ circular region. The left panel shows the radial profiles for the (NW-jet) cone (positive distances) and for the SE cone (negative distances), while the right panel shows the total ((NW-jet)+SE) bicone emission.}
 \label{fig:appb3} 
 \end{figure}

We have further compared the 4-7 keV PKS 1055+201 radial profile with the 4-7 keV \texttt{marx} model, blur=0$''$, out to 30$''$. This is shown in Fig. \ref{fig:appb3} Within the $\sim$1$\sigma$ uncertainties the PKS 1055+201 profile is consistent with the \texttt{marx} model, validating this model. However, the uncertainties are large past $\sim$7$''$.

\clearpage
\section{Comparison of the \texttt{marx} PSF model with the observed Her X-1 low-state profile}
\label{sec:appc}

\setcounter{figure}{0}
\renewcommand{\thefigure}{C\arabic{figure}}

Here we investigate how much the high signal-to-noise 4-7 keV \texttt{marx} model PSF may underestimate the wings of the PSF caused by scattering due to surface imperfections in the \textit{Chandra} mirror. The CXC performed a deep in-flight observation of the wings with ACIS-S imaging for this calibration, using the direct imaging with ACIS-S of Her X-1. The results are discussed in detail in the memorandum by \citet{gaetz2010a}\footnote{https://cxc.harvard.edu/cal/Acis/Papers/wing\_analysis\_rev1b.pdf}, which we summarize below.

Only the imaging ACIS-S observations (ObsID 3662) were used by \citet{gaetz2010a}, because both the profile extracted from the zero-th order HETG grating image of the source in low state -with low pileup, and other HETG low-pileup profiles were found to disagree significantly from the deep HRC-I observation of AR Lac, and therefore are not considered a good representation of the imaging PSF. While the reason for this is not understood, this suggests that direct imaging should be used only to study the PSF shape. For the same reason, the direct imaging of the Her X-1 profile was not normalized to the HETG profiles for comparison with the XRCF ground calibration of the PSF. A different method was then used for this normalization, based on the transfer streak (Appendix B of \citealt{gaetz2010a}). The radial extent of the pileup on the radial profile was estimated to be significant out to 8$''$-15$''$, affecting the inner core, outer core, and ``near" wings of the PSF (Appendix A of \citealt{gaetz2010a}). This estimate was based on the CCD grade migration and comparison of good and bad CCD grades in the Her X-1 ObsID 3662. Fig. A.2 from \citet{gaetz2010a} shows the radial distributions of bad and good CCD grades in Her X-1 (ObsID 3662), and their ratio. Grade migration (pileup) is observed from 15$''$ inwards, becoming increasingly severe at the center. Appendix B of \citet{gaetz2010a} gives a detailed description of the use of the readout streak to evaluate the count rate at the different energies, the evaluation and correction of the readout streak pileup ($\sim$4\%) and other corrections (Appendix C of \citealt{gaetz2010a}). The normalized data were then compared to the XRCF pre-launch calibration.

A comparison of the functional shape of the wings at different energies (see \citealt{gaetz2010a}) with the grating zero-th order image of Her X-1 in low state (ObsID 2749), can be found in the paper by \citet{gaetz2004a}, who found good functional agreement with the calibration results, once the wings were normalized to the low-state radial profile.

Fig. \ref{fig:appc1} below shows a comparison of the high signal-to-noise 4-7 keV \texttt{marx} model PSF derived from the spectrum of ObsID 2749, with the observed radial profile of the observation, both plotted in the same radial bins. Given the results of the comparison of the \texttt{marx} model with the empirical PSF from the observation of PKS 1055+201 (Appendix \ref{sec:appb}), we used an \textit{AspectBlur}=0$''$\footnote{https://cxc.cfa.harvard.edu/ciao/why/aspectblur.html} to generate the \texttt{marx} model. We normalized the model to the data at 1.75$''$, to avoid possible pileup effects (and/or discrepancy with the imaging PSF modeled by \texttt{marx}  – see \citealt{gaetz2010a}) in the core. We also compared the radial profile of ObsID 2749 with the \texttt{marx} PSF model generated with an \textit{AspectBlur}=0.2$''$. In both cases, we find that the \texttt{marx} model is in excellent agreement with the profile of ObsID 2749 in the 1$''$-5$''$ range, which is relevant for the radial profiles discussed in this paper. As shown in Appendix \ref{sec:appb}, the \texttt{marx} model is also a good representation of the PSF at radii $<$1$''$. The \texttt{marx} model underestimates the PSF wings by a factor of $\sim$6 near 10$''$. At these outer radii, the statistical uncertainties dominate the radial profiles from the data discussed in this paper.

\begin{figure}[htb]
  \centering
 \includegraphics[width=17cm]{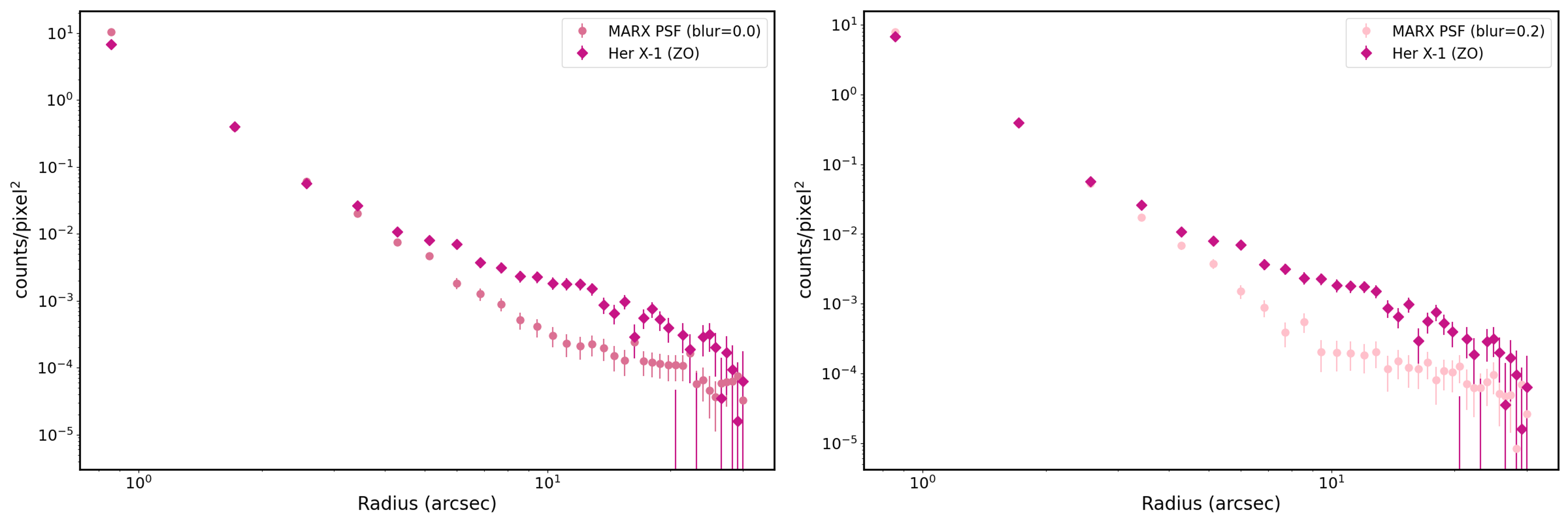}
\caption{Comparison of the high signal-to-noise 4-7 keV \texttt{marx} model PSF with the profile derived in the same energy range from the zero-order HETG observation (ObsID 2749) of Her X-1 in low state, for \textit{AspectBlur}=0$''$ and \textit{AspectBlur}=0.2$''$. Both profiles were derived with 1/8 pixel binning and normalized at 1.75$''$.}
 \label{fig:appc1} 
 \end{figure}

\clearpage
\bibliographystyle{aasjournal}
\bibliography{anna_bibliography}

\end{document}